\documentclass[10pt,twocolumn,letterpaper]{article}

\usepackage{times}
\usepackage{epsfig}
\usepackage{graphicx}
\usepackage{amsmath}
\usepackage{amssymb}
\usepackage{sidecap}
\usepackage{epstopdf}
\usepackage{multirow}

\usepackage{lineno}
\usepackage{caption}
\usepackage{subcaption}
\usepackage[strings]{underscore}
\usepackage[english]{babel}
\usepackage{sectsty}
\usepackage{titlesec}

\newsavebox\CBox
\def\textBF#1{\sbox\CBox{#1}\resizebox{\wd\CBox}{\ht\CBox}{\textbf{#1}}}

\newcommand{\eg}{{\emph{e.g.},\ }}

\usepackage[pagebackref=true,breaklinks=true,letterpaper=true,colorlinks,bookmarks=false]{hyperref}
\usepackage[top = 2.5cm, bottom = 2.50cm, left = 2.00cm, right = 2.00cm]{geometry}
\sectionfont{\fontsize{13}{18}\selectfont}
\titlelabel{\thetitle.\quad}

\begin{document}

%%%%%%%%% TITLE
%\title{\fontsize{15}{20}\selectfont \textBF{AI Benchmark: All About Deep Learning on Smartphones in 2019} \vspace{6mm}}
\title{\fontsize{15}{20}\fontseries{b}\selectfont AI Benchmark: All About Deep Learning on Smartphones in 2019 \vspace{4mm}}

\author{Andrey Ignatov\\
\small{ETH Zurich}\\
{\tt\footnotesize andrey@vision.ee.ethz.ch}
\and
Radu Timofte\\
\small{ETH Zurich}\\
{\tt\footnotesize timofter@vision.ee.ethz.ch}
\and
Andrei Kulik\\
\small{Google Research}\\
{\tt\footnotesize akulik@google.com}
\and
Seungsoo Yang\\
\small{Samsung, Inc.}\\
{\tt\footnotesize ss1.yang@samsung.com}
\and
Ke Wang\\
\small{Huawei, Inc.}\\
{\tt\footnotesize michael.wangke@huawei.com}
\and
Felix Baum\\
\small{Qualcomm, Inc.}\\
{\tt\footnotesize fbaum@qti.qualcomm.com}
\and
Max Wu\\
\small{MediaTek, Inc.}\\
{\tt\footnotesize max.wu@mediatek.com}
\and
Lirong Xu\\
\small{Unisoc, Inc.}\\
{\tt\footnotesize lirong.Xu@unisoc.com}
\and
Luc Van Gool\thanks{We also thank Oli Gaymond (ogaymond@google.com), Google Inc., for writing and editing section 3.1 of this paper.}\\
\small{ETH Zurich}\\
{\tt\footnotesize vangool@vision.ee.ethz.ch}
\vspace{0.2mm}
}

\date{}
\maketitle

%%%%%%%%% ABSTRACT
\begin{abstract}
\textit{The performance of mobile AI accelerators has been evolving rapidly in the past two years, nearly doubling with each new generation of SoCs. The current 4th generation of mobile NPUs is already approaching the results of CUDA-compatible Nvidia graphics cards presented not long ago, which together with the increased capabilities of mobile deep learning frameworks makes it possible to run complex and deep AI models on mobile devices. In this paper, we evaluate the performance and compare the results of all chipsets from Qualcomm, HiSilicon, Samsung, MediaTek and Unisoc that are providing hardware acceleration for AI inference. We also discuss the recent changes in the Android ML pipeline and provide an overview of the deployment of deep learning models on mobile devices. All numerical results provided in this paper can be found and are regularly updated on the official project website~\footnote{\url{http://ai-benchmark.com}}.}
\end{abstract}

%%%%%%%%% BODY TEXT
\section{Introduction}
\label{sec:introduction}

Over the past years, deep learning and AI became one of the key trends in the mobile industry. This was a natural fit, as from the end of the 90s mobile devices were getting equipped with more and more software for intelligent data processing~-- face and eyes detection~\cite{hadid2007face}, eye tracking~\cite{miluzzo2010eyephone}, voice recognition~\cite{matsunaga2005universal}, barcode scanning~\cite{von2010evaluation}, accelerometer-based gesture recognition~\cite{liu2009uwave,niezen2008gesture}, predictive text recognition~\cite{silfverberg2000predicting}, handwritten text recognition~\cite{anquetil2002integration}, OCR~\cite{koga2005camera}, etc. At the beginning, all proposed methods were mainly based on manually designed features and very compact models as they were running at best on devices with a single-core 600 MHz Arm CPU and 8-128 MB of RAM. The situation changed after 2010, when mobile devices started to get multi-core processors, as well as powerful GPUs, DSPs and NPUs, well suitable for machine and deep learning tasks. At the same time, there was a fast development of the deep learning field, with numerous novel approaches and models that were achieving a fundamentally new level of performance for many practical tasks, such as image classification, photo and speech processing, neural language understanding, etc. Since then, the previously used hand-crafted solutions were gradually replaced by considerably more powerful and efficient deep learning techniques, bringing us to the current state of AI applications on smartphones.

Nowadays, various deep learning models can be found in nearly any mobile device. Among the most popular tasks are different computer vision problems like image classification~\cite{krizhevsky2012imagenet,szegedy2016rethinking,howard2017mobilenets}, image enhancement~\cite{ignatov2017dslr,ignatov2017wespe,ignatov2018pirm,ignatov2019ntire}, image super-resolution~\cite{dong2016image,ledig2017photo,timofte2018ntire}, bokeh simulation~\cite{wadhwa2018synthetic}, object tracking~\cite{wu2015object,huang2017speed}, optical character recognition~\cite{netzer2011reading}, face detection and recognition~\cite{li2015convolutional,schroff2015facenet}, augmented reality~\cite{alhaija2017augmented,detone2016deep}, etc. Another important group of tasks running on mobile devices is related to various NLP (Natural Language Processing) problems, such as natural language translation~\cite{sutskever2014sequence,bahdanau2014neural}, sentence completion~\cite{mikolov2013efficient,hu2014convolutional}, sentence sentiment analysis~\cite{socher2013recursive,severyn2015twitter,ignatov2017decision}, voice assistants~\cite{emmett2013voice} and interactive chatbots~\cite{serban2017deep}. Additionally, many tasks deal with time series processing, \eg human activity recognition~\cite{kwapisz2011activity,ignatov2018real}, gesture recognition~\cite{ordonez2016deep}, sleep monitoring~\cite{sathyanarayana2016sleep}, adaptive power management~\cite{mannor2006machine,liu2010adaptive}, music tracking~\cite{wang2006shazam} and classification~\cite{sigtia2014improved}. Lots of machine and deep learning algorithms are also integrated directly into smartphones firmware and used as auxiliary methods for estimating various parameters and for intelligent data processing.

\begin{figure*}[t!]
\centering
\resizebox{1.0\linewidth}{!}
{
\includegraphics[width=1.0\linewidth]{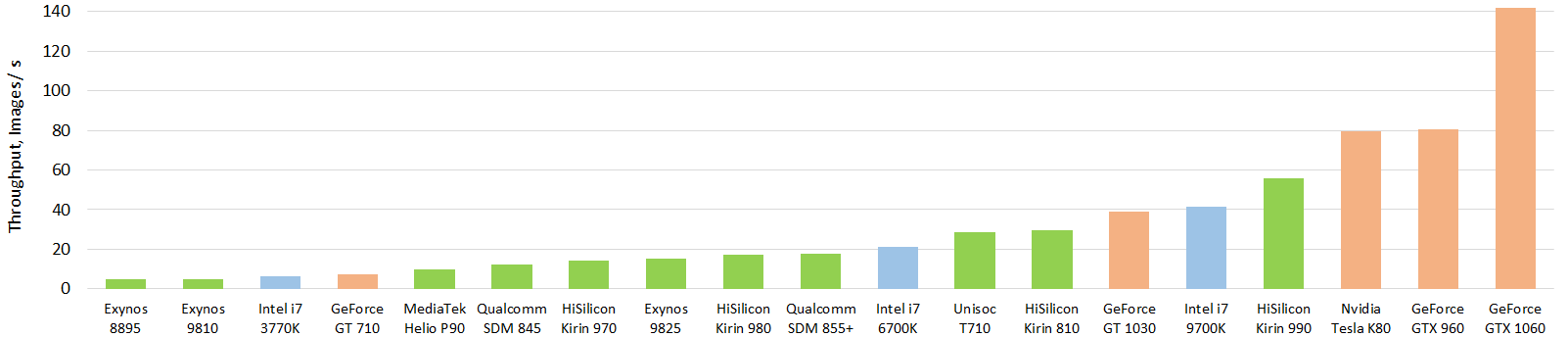}
}
\vspace{0mm}
\caption{\small{Performance evolution of mobile AI accelerators: image throughput for the float Inception-V3 model. Mobile devices were running the FP16 model using TensorFlow Lite and NNAPI. Acceleration on Intel CPUs was achieved using the Intel MKL-DNN library~\cite{MKLDNN2019}, on Nvidia GPUs~-- with CUDA~\cite{CUDA2019} and cuDNN~\cite{chetlur2014cudnn}. The results on Intel and Nvidia hardware were obtained using the standard TensorFlow library~\cite{abadi2016tensorflow} running the FP32 model with a batch size of 20 (the FP16 format is currently not supported by these CPUs / GPUs). Note that the Inception-V3 is a relatively small network, and for bigger models the advantage of Nvidia GPUs over other silicon might be larger.}}
\label{fig:ai_evolution}
\vspace{-0.4mm}
\end{figure*}

While running many state-of-the-art deep learning models on smartphones was initially a challenge as they are usually not optimized for mobile inference, the last few years have radically changed this situation. Presented back in 2015, TensorFlow Mobile~\cite{TensorFlowMobile2018} was the first official library allowing to run standard AI models on mobile devices without any special modification or conversion, though also without any hardware acceleration, i.e.~on CPU only. In 2017, the latter limitation was lifted by the TensorFlow Lite (TFLite)~\cite{TensorFlowLite2018} framework that dropped support for many vital deep learning operations, but offered a significantly reduced binary size and kernels optimized for on-device inference. This library also got support for the Android Neural Networks API (NNAPI)~\cite{NNAPI2018}, introduced the same year and allowing for the access to the device's AI hardware acceleration resources directly through the Android operating system. This was an important milestone as a full-fledged mobile ML pipeline was finally established: training, exporting and running the resulting models on mobile devices became possible within one standard deep learning library, without using specialized vendors tools or SDKs. At first, however, this approach had also numerous flaws related to NNAPI and TensorFlow Lite themselves, thus making it impractical for many use cases. The most notable issues were the lack of valid NNAPI drivers in the majority of Android devices (only 4 commercial models featured them as of September 2018~\cite{AIBenchmarkSept2018}), and the lack of support for many popular ML models by TFLite. These two issues were largely resolved during the past year. Since the spring of 2019, nearly all new devices with Qualcomm, HiSilicon, Samsung and MediaTek systems on a chip (SoCs) and with dedicated AI hardware are shipped with NNAPI drivers allowing to run ML workloads on embedded AI accelerators. In Android 10, the Neural Networks API was upgraded to version 1.2 that implements 60 new ops~\cite{NNAPI122019} and extends the range of supported models. Many of these ops were also added to TensorFlow Lite starting from builds 1.14 and 1.15. Another important change was the introduction of TFLite delegates~\cite{TFLiteDelegates2019}. These delegates can be written directly by hardware vendors and then used for accelerating AI inference on devices with outdated or absent NNAPI drivers. A universal delegate for accelerating deep learning models on mobile GPUs (based on OpenGL ES, OpenCL or Metal) was already released by Google earlier this year~\cite{lee2019device}. All these changes build the foundation for a new mobile AI infrastructure tightly connected with the standard machine learning (ML) environment, thus making the deployment of machine learning models on smartphones easy and convenient. The above changes will be described in detail in Section~\ref{sec:deep_learning_on_smartphones}.

The latest generation of mid-range and high-end mobile SoCs comes with AI hardware, the performance of which is getting close to the results of desktop CUDA-enabled Nvidia GPUs released in the past years. In this paper, we present and analyze performance results of all generations of mobile AI accelerators from Qualcomm, HiSilicon, Samsung, MediaTek and Unisoc, starting from the first mobile NPUs released back in 2017. We compare against the results obtained with desktop GPUs and CPUs, thus assessing performance of mobile vs. conventional machine learning silicon. To do this, we use a professional AI Benchmark application~\cite{ignatov2018ai} consisting of 21 deep learning tests and measuring more than 50 different aspects of AI performance, including the speed, accuracy, initialization time, stability, etc. The benchmark was significantly updated since previous year to reflect the latest changes in the ML ecosystem. These updates are described in Section~\ref{sec:ai_benchmark_3}. Finally, we provide an overview of the performance, functionality and usage of Android ML inference tools and libraries, and show the performance of more than 200 Android devices and 100 mobile SoCs collected in-the-wild with the AI Benchmark application.

The rest of the paper is arranged as follows. In Section~\ref{sec:hardware_acceleration} we describe the hardware acceleration resources available on the main chipset platforms and programming interfaces to access them. Section~\ref{sec:deep_learning_on_smartphones} gives an overview of the latest changes in the mobile machine learning ecosystem. Section~\ref{sec:ai_benchmark_3} provides a detailed description of the recent modifications in our AI Benchmark architecture, its programming implementation and deep learning tests. Section~\ref{sec:benchmark_results} shows the experimental performance results for various mobile devices and chipsets, and compares them to the performance of desktop CPUs and GPUs. Section~\ref{sec:discussion} analyzes the results. Finally, Section~\ref{sec:conclusion} concludes the paper.

\vspace{-2mm}
\section{Hardware Acceleration}
\label{sec:hardware_acceleration}

Though many deep learning algorithms were presented back in the 1990s ~\cite{lecun1989backpropagation,lecun1998gradient,hochreiter1997long}, the lack of appropriate (and affordable) hardware to train such models prevented them from being extensively used by the research community till 2009, when it became possible to effectively accelerate their training with general-purpose consumer GPUs~\cite{raina2009large}. With the introduction of Max-Pooling CNNs~\cite{ciresan2011flexible,nagi2011max} and AlexNet~\cite{krizhevsky2012imagenet} in 2011-2012 and the subsequent success of deep learning in many practical tasks, it was only a matter of time before deep neural networks would be run on mobile devices. Compared to simple statistical methods previously deployed on smartphones, deep learning models required huge computational resources and thus running them on Arm CPUs was nearly infeasible from both the performance and power efficiency perspective. The first attempts to accelerate AI models on mobile GPUs and DSPs were made in 2015 by Qualcomm~\cite{Qualcomm2015Zeroth}, Arm~\cite{Mali2015OpenCL} and other SoC vendors, though at the beginning mainly by adapting deep learning models to the existing hardware. Specialized AI silicon started to appear in mobile SoCs with the release of the Snapdragon 820 / 835 with the Hexagon V6 68x DSP series optimized for AI inference, the Kirin 970 with a dedicated NPU unit designed by Cambricon, the Exynos 8895 with a separate Vision Processing Unit, MediaTek Helio P60 with AI Processing Unit, and the Google Pixel 2 with a standalone Pixel Visual Core. The performance of mobile AI accelerators has been evolving extremely rapidly in the past three years (Fig.~\ref{fig:ai_evolution}), coming ever closer to the results of desktop hardware. We can now distinguish four generations of mobile SoCs based on their AI performance, capabilities and release date:

\smallskip

\textBF{Generation 1:}\, All legacy chipsets that can not provide AI acceleration through the Android operating system, but still can be used to accelerate machine learning inference with special SDKs or GPU-based libraries. All Qualcomm SoCs with Hexagon 682 DSP and below, and the majority of chipsets from HiSilicon, Samsung and MediaTek belong to this category. It is worth mentioning that nearly all computer vision models are largely based on vector and matrix multiplications, and thus can technically run on almost any mobile GPU supporting OpenGL ES or OpenCL. Yet, this approach might actually lead to notable performance degradation on many SoCs with low-end or old-gen GPUs.

\smallskip

\begin{figure}[t!]
\centering
\resizebox{1.0\linewidth}{!}
{
\includegraphics[width=1.0\linewidth]{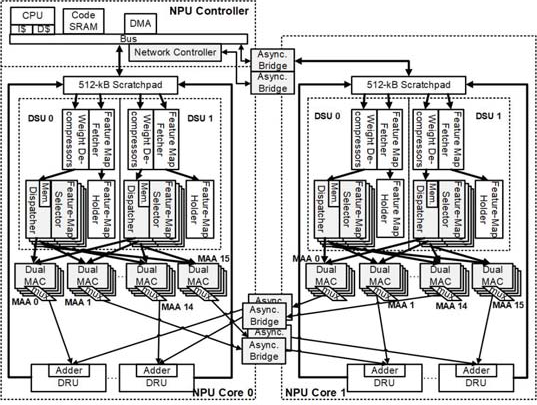}
}
\vspace{-2mm}
\caption{\small{The overall architecture of the Exynos 9820 NPU~\cite{song20197}.}}
\label{fig:ExynosNPU}
\vspace{-2.2mm}
\end{figure}

\textBF{Generation 2:}\, Mobile SoCs supporting Android NNAPI and released after 2017. They might provide acceleration for only one type of models (float or quantized) and are typical for the AI performance in 2018.

\smallskip

\noindent \hspace{0.2mm} $\bullet$ Qualcomm: Snapdragon 845 (Hex. 685 + Adreno~630);

\hspace{16.2mm} Snapdragon 710 (Hexagon 685);

\hspace{16.2mm} Snapdragon 670 (Hexagon 685);

\smallskip

\noindent \hspace{0.2mm} $\bullet$ HiSilicon: \hspace{0.9mm} Kirin 970 (NPU, Cambricon);

\smallskip

\noindent \hspace{0.2mm} $\bullet$ Samsung: \hspace{1.6mm} Exynos 9810 (Mali-G72 MP18);

\hspace{16.2mm} Exynos 9610 (Mali-G72 MP3);

\hspace{16.2mm} Exynos 9609 (Mali-G72 MP3);

\smallskip

\noindent \hspace{0.2mm} $\bullet$ MediaTek: \hspace{0.2mm} Helio P70 (APU 1.0 + Mali-G72 MP3);

\hspace{16.2mm} Helio P60 (APU 1.0 + Mali-G72 MP3);

\hspace{16.2mm} Helio P65 (Mali-G52 MP2).

\smallskip

\textBF{Generation 3.}\, Mobile SoCs supporting Android NNAPI and released after 2018. They provide hardware acceleration for all model types and their AI performance is typical for the corresponding SoC segment in 2019.

\smallskip

\noindent \hspace{0.2mm} $\bullet$ Qualcomm: Snapdragon 855+ (Hex. 690 + Adreno~640);

\hspace{16.2mm} Snapdragon 855 (Hex. 690 + Adreno~640);

\hspace{16.2mm} Snapdragon 730 (Hex. 688 + Adreno~618);

\hspace{16.2mm} Snapdragon 675 (Hex. 685 + Adreno~612);

\hspace{16.2mm} Snapdragon 665 (Hex. 686 + Adreno~610);

\smallskip

\noindent \hspace{0.2mm} $\bullet$ HiSilicon: \hspace{0.9mm} Kirin 980 (NPU$\times$2, Cambricon);

\smallskip

\noindent \hspace{0.2mm} $\bullet$ Samsung: \hspace{1.6mm} Exynos 9825 (NPU +  Mali-G76 MP12);

\hspace{16.2mm} Exynos 9820 (NPU + Mali-G76 MP12);

\smallskip

\noindent \hspace{0.2mm} $\bullet$ MediaTek: \hspace{0.2mm} Helio P90 (APU 2.0);

\hspace{16.2mm} Helio G90 (APU 1.0 + Mali-G76 MP4).

\smallskip

\textBF{Generation 4:}\, Recently presented chipsets with next-generation AI accelerators (Fig.~\ref{fig:ai_evolution}). Right now, only the HiSilicon Kirin 990, HiSilicon Kirin 810 and Unisoc Tiger T710 SoCs belong to this category. Many more chipsets from other vendors will come by the end of this year.

\smallskip

Below, we provide a detailed description of the mobile platforms and related SDKs released in the past year. More information about SoCs with AI acceleration support that were introduced earlier, can be found in our previous paper~\cite{ignatov2018ai}.

\subsection{Samsung chipsets / EDEN SDK}

The Exynos 9820 was the first Samsung SoC to get an NPU technically compatible with Android NNAPI, its drivers will be released after Android Q upgrade. This chipset contains two custom Mongoose M4 CPU cores, two Cortex-A75, four Cortex-A55 cores and Mali-G76 MP12 graphics. The NPU of the Exynos 9820 supports only quantized inference and consists of the controller and two cores (Fig.~\ref{fig:ExynosNPU}) having 1024 multiply-accumulate (MAC) units~\cite{song20197}.
The NPU controller has a CPU, a direct memory access (DMA) unit, code SRAM and a network controller. The CPU is communicating with the host system of the SoC and defines the network scale for the network controller. The controller automatically configures all modules in the two cores and traverses the network. To use the external memory bandwidth and the scratchpads efficiently, the weights of the network are compressed, and the network compiler additionally partitions the network into sub-networks and performs the traversal over multiple network layers. The DMA unit manages the compressed weights and feature maps in each of the 512KB scratchpads of the cores. When running the computations, the NPU can also skip weights that are zero to improve convolution efficiency. A much more detailed description of the Exynos NPU can be found in~\cite{song20197}. We strongly recommend reading this article for everyone interested in the general functioning of NPUs as it provides an excellent overview on all network / data processing stages and possible bottlenecks.

\begin{figure}[t!]
\centering
\resizebox{1.0\linewidth}{!}
{
\includegraphics[width=1.0\linewidth]{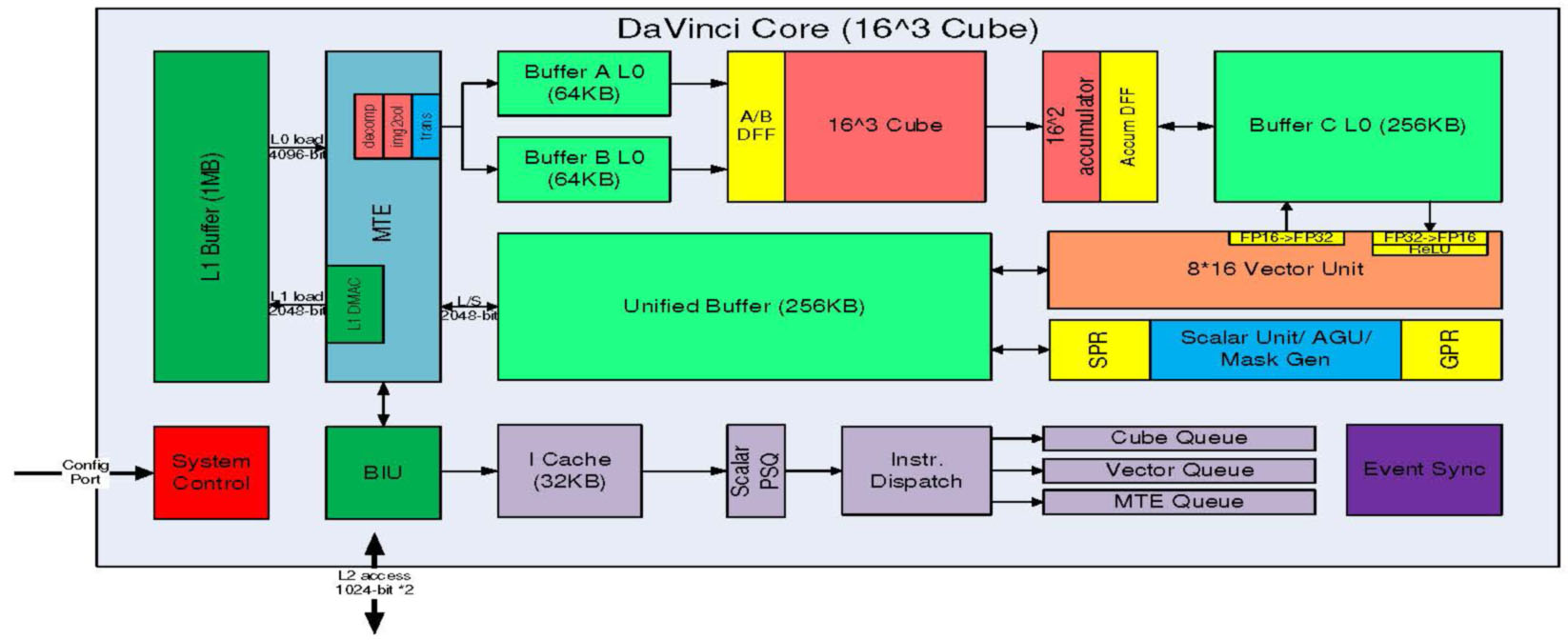}
}
\vspace{-1.2mm}
\caption{\small{A general architecture of the Huawei's DaVinci Core.}}
\label{fig:daVinci}
\vspace{-2mm}
\end{figure}

The Exynos 9820's NPU occupies 5.5mm$^2$, is fabricated in 8nm CMOS technology and operates at 67-933 MHz clock frequency. The performance of the NPU heavily depends on the kernel sizes and the fraction of zero weights. For kernels of size 5$\times$5, it achieves the performance of 2.1 TOPS and 6.9 TOPS for 0\% and 75\% zero-weights, respectively; the energy efficiency in these two cases is 3.6 TOPS/W and \mbox{11.5 TOPS/W}. For the Inception-V3 model, the energy efficiency lies between \mbox{2 TOPS/W} and \mbox{3.4 TOPS/W} depending on network sparsity~\cite{song20197}.

The other two Samsung SoCs that support Android NNAPI are the Exynos 9609 / 9610, though they are relying on the Mali-G72 MP3 GPU and Arm NN drivers~\cite{ArmNN2018} to accelerate AI models. As to the Exynos 9825 presented together with the latest Note10 smartphone series, this is a slightly overclocked version of the Exynos 9820 produced in 7nm technology, with the same NPU design.

%This year, Samsung released the SDK~\cite{EDEN2019} for the acceleration of AI computations on Samsung devices with GPUs and NPUs. This SDK is now mainly based on the Arm NN~\cite{ArmNN2018} and Qualcomm SNPE~\cite{SNPE2018} libraries, and is currently still under development, supporting only Caffe and TensorFlow models (with the latter now running on CPU only).

This year, Samsung announced the Exynos Deep Neural Network (EDEN) SDK that provides the NPU, GPU and CPU acceleration for deep learning models and exploits the data and model parallelism. It consists of the model conversion tool, the NPU compiler and the customized TFLite generator and is available as a desktop tool plus runtimes for Android and Linux. The EDEN runtime provides APIs for initialization, opening / closing the model and its execution with various configurations. Unfortunately, it is not publicly available yet.

\subsection{HiSilicon chipsets / HiAI SDK}

\begin{figure}[t!]
\centering
\resizebox{0.95\linewidth}{!}
{
\includegraphics[width=1.0\linewidth]{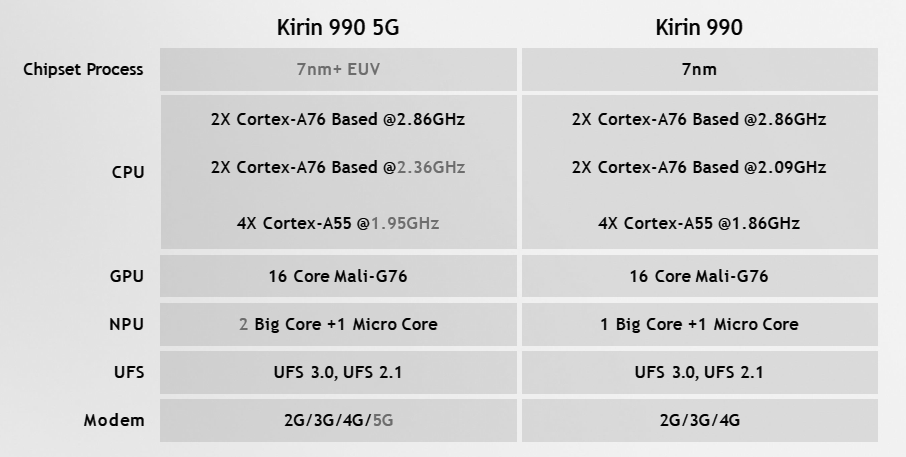}
}
\vspace{2.6mm}
\caption{\small{SoC components integrated into the Kirin 990 chips.}}
\label{fig:Kirin990}
\vspace{-2mm}
\end{figure}

While the Kirin 970 / 980 SoCs were using NPUs originally designed by Cambricon, this year Huawei switched to its in-house developed Da Vinci architecture (Fig.~\ref{fig:daVinci}), powering the Ascend series of AI accelerators and using a 3D Cube computing engine to accelerate matrix computations. The first SoC with Da Vinci NPU was a mid-range Kirin 810 incorporating two Cortex-A76 and six Cortex-A55 CPU cores with Mali-G52 MP6 GPU. A significantly enlarged AI accelerator appeared later in the Kirin 990 5G chip having four Cortex-A76, four Cortex-A55 CPUs and Mali-G76 MP16 graphics. This SoC features a triple-core \mbox{Da Vinci} NPU containing two large (Da Vinci Lite) cores for heavy computing scenarios and one little (Da Vinci Tiny) core for low-power AI computations. According to Huawei, the little core is up to 24 times more power efficient than the large one when running face recognition models. Besides that, a simplified version of the Kirin 990 (without ``5G'' prefix) with a dual-core NPU (one large + one small core) was also presented and should not be confused with the standard version (Fig.~\ref{fig:Kirin990}).

In the late 2018, Huawei launched the HiAI 2.0 SDK with added support for the Kirin 980 chipset and new deep learning ops. Huawei has also released the IDE tool and Android Studio plug-in, providing development toolsets for running deep learning models with the HiAI Engine. With the recent update of HiAI, it supports more than 300 deep learning ops and the latest Kirin 810 / 990 (5G) SoCs.

\subsection{Qualcomm chipsets / SNPE SDK}

As before, Qualcomm is relying on its AI Engine (consisting of the Hexagon DSP, Adreno GPU and Kryo CPU cores) for the acceleration of AI inference. In all Qualcomm SoCs supporting Android NNAPI, the Adreno GPU is used for floating-point deep learning models, while the Hexagon DSP is responsible for quantized inference. It should be noted that though the Hexagon 68x/69x chips are still marketed as DSPs, their architecture was optimized for deep learning workloads and they include dedicated AI silicon such as tensor accelerator units, thus not being that different from NPUs and TPUs proposed by other vendors. The only major weakness of the Hexagon DSPs is the lack of support for floating-point models (same as in the Google Pixel TPU, MediaTek APU 1.0 and Exynos NPU), thus the latter are delegated to Adreno GPUs.

\begin{figure}[t!]
\centering
\resizebox{0.95\linewidth}{!}
{
\includegraphics[width=0.715\linewidth]{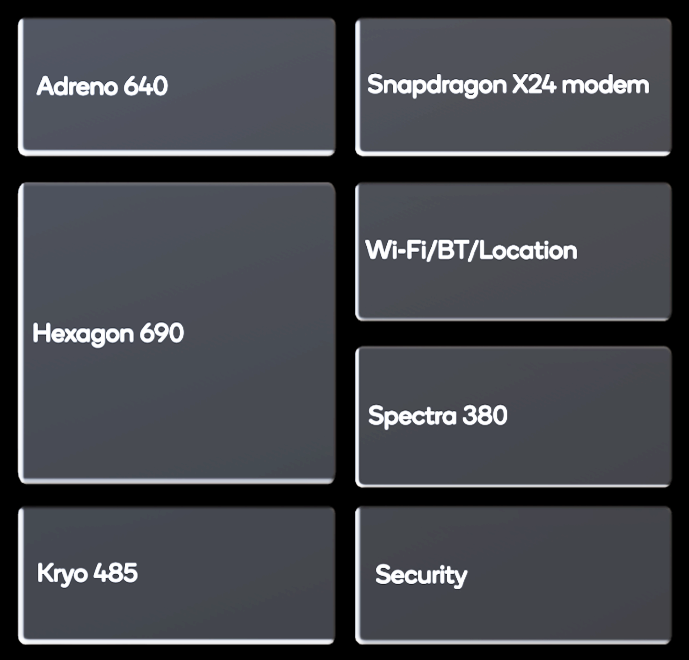}\hspace{10.2mm}
\includegraphics[width=0.785\linewidth]{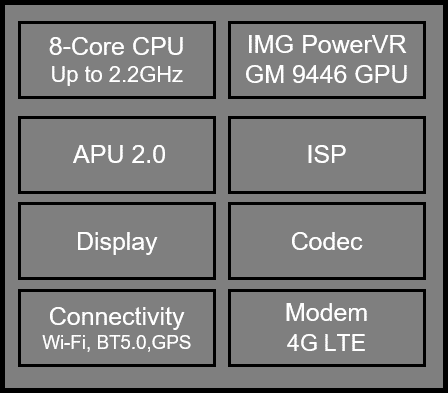}
}
\vspace{2.6mm}
\caption{\small{Qualcomm Snapdragon 855 (left) and MediaTek Helio P90 (right) block diagrams.}}
\label{fig:SoC_diagrams}
\vspace{-2mm}
\end{figure}

At the end of 2018, Qualcomm announced its flagship SoC, the Snapdragon 855, containing eight custom Kryo 485 CPU cores (three clusters functioning at different frequencies, Cortex-A76 derived), an Adreno 640 GPU and Hexagon 690 DSP (Fig.~\ref{fig:SoC_diagrams}). Compared to the Hexagon 685 used in the SDM845, the new DSP got a 1024-bit SIMD with  double the number of pipelines and an additional tensor accelerator unit. Its GPU was also upgraded from the previous generation, getting twice more ALUs and an expected performance increase of 20\% compared to the Adreno 630. The Snapdragon 855 Plus, released in July 2019, is an overclocked version of the standard SDM855 SoC, with the same DSP and GPU working at higher frequencies. The other three mid-range SoCs introduced in the past year (Snapdragon 730, 665 and 675) include the Hexagon 688, 686 and 685 DSPs, respectively (the first two are derivatives of the Hexagon 685). All the above mentioned SoCs support Android NNAPI 1.1 and provide acceleration for both float and quantized models. According to Qualcomm, all NNAPI-compliant chipsets (Snapdragon 855, 845, 730, 710, 675, 670 and 665) will get support for NNAPI 1.2 in Android Q.

Qualcomm's Neural Processing SDK (SNPE)~\cite{SNPE2018} also went through several updates in the past year. It currently offers Android and Linux runtimes for neural network model execution, APIs for controlling loading / execution / scheduling on the runtimes, desktop tools for model conversion and a performance benchmark for bottleneck identification. It currently supports the Caffe, Caffe2, ONNX and TensorFlow machine learning frameworks.

\subsection{MediaTek chipsets / NeuroPilot SDK}

One of the key releases from MediaTek in the past year was the Helio P90 with a new AI Processing Unit (APU 2.0) that can generate a computational power of up to 1.1 TMACs / second (4 times higher than the previous Helio P60 / P70 series). The SoC, manufactured with a 12nm process, combines a pair of Arm Cortex-A75 and six Cortex-A55 CPU cores with the IMG PowerVR GM 9446 GPU and dual-Channel LPDDR4x RAM up to 1866MHz. The design of the APU was optimized for operations intensively used in deep neural networks. First of all, its parallel processing engines are capable of accelerating heavy computing operations, such as convolutions, fully connected layers, activation functions, 2D operations (\eg pooling or bilinear interpolation) and other tensor manipulations. The task control system and data buffer were designed to minimize memory bandwidth usage and to maximize data reuse and the utilization rate of processing engines. Finally, the APU is supporting all popular inference modes, including FP16, INT16 and INT8, allowing to run all common AI models with hardware acceleration. Taking face detection as an example, the APU can run up to 20 times faster and reduce the power consumption by 55 times compared to the Helio's CPU.

\begin{figure}[t!]
\centering
\resizebox{1.0\linewidth}{!}
{
\includegraphics[width=1.0\linewidth]{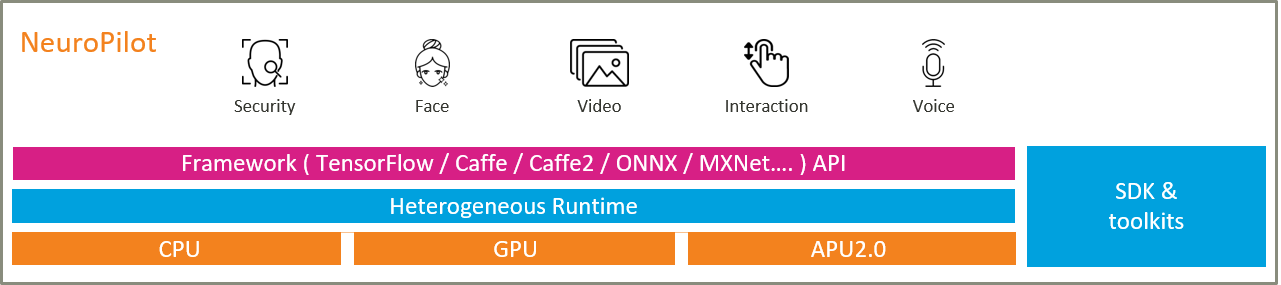}
}
\vspace{-1.8mm}
\caption{\small{Schematic representation of MediaTek NeuroPilot SDK.}}
\label{fig:NeuroPilotSdk}
\vspace{-2mm}
\end{figure}

As to other MediaTek chipsets presented this year, the Helio G90 and the Helio P65 are also providing hardware acceleration for float and quantized AI models. The former uses a separate APU (1st gen.) with a similar architecture as the one in the Helio P60 / P70 chipsets~\cite{ignatov2018ai}. The Helio P65 does not have a dedicated APU module and is running all models on a Mali-G52 MP2 GPU.

Together with the Helio P90, MediaTek has also launched the NeuroPilot v2.0 SDK (Fig.~\ref{fig:NeuroPilotSdk}). In its second version, NeuroPilot supports automatic network quantization and pruning. The SDK's APU drivers support FP16/INT16/INT8 data types, while CPU and GPU drivers can be used for some custom ops and FP32/FP16 models. The NeuroPilot SDK was designed to take advantage of MediaTek's heterogeneous hardware, by assigning the workloads to the most suitable processor and concurrently utilizing all available computing resources for the best performance and energy efficiency. The SDK is supporting only MediaTek NeuroPilot-compatible chipsets across products such as smartphones and TVs. At its presentation of the Helio P90, MediaTek demonstrated that NeuroPilot v2.0 allows for the real-time implementation of many AI applications (e.g. multi-person pose tracking, 3D pose tracking, multiple object identification, AR / MR, semantic segmentation, scene identification and image enhancement).

\subsection{Unisoc chipsets / UNIAI SDK}

Unisoc is a Chinese fabless semiconductor company (formerly known as Spreadtrum) founded in 2001. The company originally produced chips for GSM handsets and was mainly known in China, though starting from 2010-2011 it began to expand its business to the global market. Unisoc's first smartphone SoCs (SC8805G and SC6810) appeared in  entry-level Android devices in 2011 and were featuring an ARM-9 600MHz processor and 2D graphics. With the introduction of the quad-core Cortex-A7 based SC773x, SC883x and SC983x SoC series, Unisoc chipsets became used in many low-end, globally shipped Android devices. The performance of Unisoc's budget chips was notably improved in the SC9863 SoC and in the Tiger T310 platform released earlier this year. To target the mid-range segment, Unisoc introduced the Tiger T710 SoC platform with four Cortex-A75 + four Cortex-A55 CPU cores and IMG PowerVR GM 9446 graphics. This is the first chipset from Unisoc to feature a dedicated NPU module for the acceleration of AI computations.
The NPU of the T710 consists of two different computing accelerator cores: one for integer models supporting the INT4, INT8 and INT16 formats and providing a peak performance of 3.2 TOPS for INT8, and the other for FP16 models with 0.5 TFLOPS performance. The two cores can either accelerate different AI tasks at the same time, or accelerate the task with one of them, while the second core can be completely shut down to reduce the overall power consumption of the SoC. The Tiger T710 supports Android NNAPI and implements Android NN Unosic HIDL services supporting INT8 / FP16 models. The overall energy efficiency of the T710's NPU is greater than or equal to 2.5 TOPS/W depending on the scenario.

\begin{figure}[t!]
\centering
\resizebox{0.75\linewidth}{!}
{
\includegraphics[width=1.0\linewidth]{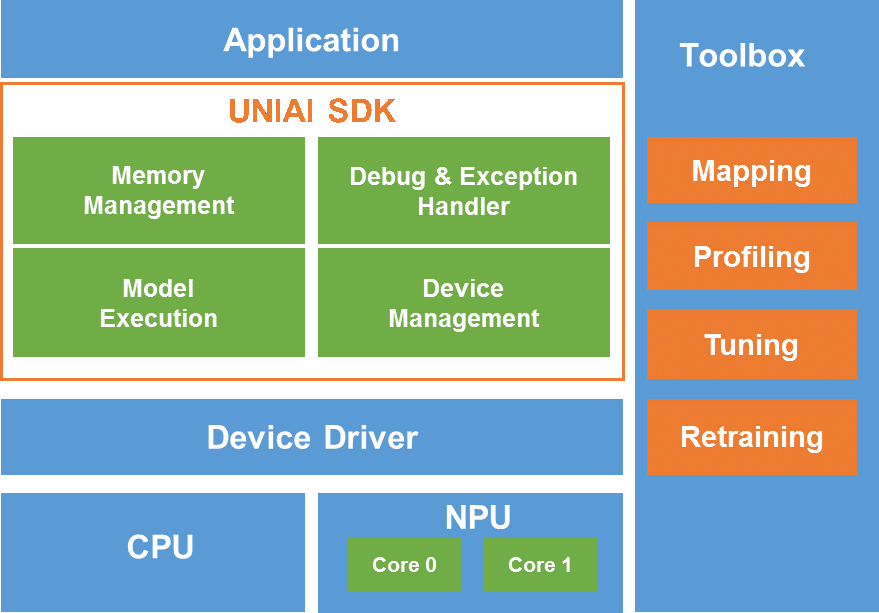}
}
\vspace{2.6mm}
\caption{\small{Schematic representation of Unisoc UNIAI SDK.}}
\label{fig:UNIAI}
\vspace{-2mm}
\end{figure}

Unisoc has also developed the UNIAI SDK~\ref{fig:UNIAI} that consists of two parts: the off-line model conversion tool that can compile the trained model into a file that can be executed on NPU; and the off-line model API and runtime used to load and execute the compiled model. The off-line model conversion tool supports several neural network framework formats, including Tensorflow, Tensorflow Lite, Caffe and ONNX. To improve the flexibility, the NPU Core also includes units that can be programmed to support user defined ops, making it possible to run the entire model with such ops on NPU and thus significantly decreasing runtime.

\subsection{Google Pixel 3 / Pixel Visual Core}

As for the Pixel 2 series, the third generation of Google phones contains a separate tensor processing unit (Pixel Visual Core) capable of accelerating deep learning ops. This TPU did not undergo significant design changes compared to the previous version. Despite Google's initial statement~\cite{Pixel2PressRelease}, neither SDK nor NNAPI drivers were or will be released for this TPU series, making it inaccessible to anyone except Google. Therefore, its importance for deep learning developers is limited. In the Pixel phones, it is used for a few tasks related to HDR photography and real-time sensor data processing.

%\subsection{Arm NPUs / Arm NN}

\section{Deep Learning on Smartphones}
\label{sec:deep_learning_on_smartphones}

In a preceding paper~(\cite{ignatov2018ai}, Section~3), we described the state of the deep learning mobile ecosystem as of September 2018. The changes in the past year were along the line of expectations. The TensorFlow Mobile~\cite{TensorFlowMobile2018} framework was completely deprecated by Google in favor of TensorFlow Lite that got a significantly improved CPU backend and support for many new ops. Yet, TFLite is still lacking some vital deep learning operators, especially those used in many NLP models. Therefore, TensorFlow Mobile remains relevant for complex architectures. Another recently added option for unsupported models is to use the TensorFlow Lite plugin containing standard TensorFlow operators~\cite{TFLiteWithTF2019} that are not yet added to TFLite. That said, the size of this plugin (40MB) is even larger than the size of the TensorFlow Mobile library (20MB). As to the Caffe2 / PyTorch libraries, while some unofficial Android ports appeared in the past 12 months~\cite{PyTorchLite2019,PyTorchPort2019}, there is still no official support for Android (except for 2 two-year old camera demos~\cite{PyTorch2018Stye,PyTorch2018AICamera}), thus making it not that interesting for regular developers.

Though some TensorFlow Lite issues mentioned last year~\cite{ignatov2018ai} were solved in its current releases, we still recommend using it with great precaution. For instance, in its latest official build (1.14), the interaction with NNAPI was completely broken, leading to enormous losses and random outputs during the first two inferences. This issue can be solved by replacing the \textit{setUseNNAPI} method with a stand-alone NNAPI delegate present in the TFLite-GPU delegate library~\cite{TFLite2019GPU}. Another problem present in the nightly builds is a significantly increased RAM consumption for some models (\eg SRCNN, Inception-ResNet-V1, VGG-19), making them crashing even on devices with 4GB+ of RAM. While these issues should be solved in the next official TFLite release (1.15), we suggest developers to extensively test their models on all available devices with each change of TFLite build. Another recommended option is to move to custom TensorFlow Lite delegates from SoC vendors that allow to omit such problems and potentially achieve even better results on their hardware.

The other two major changes in the Android deep learning ecosystem were the introduction of TensorFlow Lite delegates and Neural Networks API 1.2. We describe them in detail below.

\subsection{Android NNAPI 1.2}

The latest version of NN API provides access to 56 new operators, significantly expanding the range of models that can be supported for hardware acceleration. In addition the range of supported data types has increased, bringing support for per-axis quantization for weights and IEEE Float 16. This broader support for data types enables developers and hardware makers to determine the most performant options for their specific model needs.

A significant addition to the API surface is the ability to query the underlying hardware accelerators at runtime and specify explicitly where to run the model. This enables use cases where the developer wants to limit contention between resources, for example an Augmented Reality developer may choose to ensure the GPU is free for visual processing requirements by directing their ML workloads to an alternative accelerator if available.

Neural Networks API 1.2 introduces the concept of burst executions. Burst executions are a sequence of executions of the same prepared model that occur in rapid succession, such as those operating on frames of a camera capture or successive audio samples. A burst object is used to control a set of burst executions, and to preserve resources between executions, enabling executions to have lower overhead.
From Android 10, NNAPI provides functions to support caching of compilation artifacts, which reduces the time used for compilation when an application starts. Using this caching functionality, the driver does not need to manage or clean up the cached files.
Neural Networks API (NNAPI) vendor extensions, introduced in Android 10, are collections of vendor-defined operations and data types. On devices running NN HAL 1.2 or higher, drivers can provide custom hardware-accelerated operations by supporting corresponding vendor extensions. Vendor extensions do not modify the behavior of existing operations. Vendor extensions provide a more structured alternative to OEM operation and data types, which were deprecated in Android 10.

\subsection{TensorFlow Lite Delegates}

In the latest releases, TensorFlow Lite provides APIs for delegating the execution of neural network sub-graphs to external libraries (called delegates)~\cite{TFLiteDelegates2019}.
Given a neural network model, TFLite first checks what operators in the model can be executed with the provided delegate.
Then TFLite partitions the graph into several sub-graphs, substituting the supported by the delegate sub-graphs with virtual ``delegate nodes''~\cite{lee2019device}.
From that point, the delegate is responsible for executing all sub-graphs in the corresponding nodes.
Unsupported operators are by default computed by the CPU, though this might significantly increase the inference time as there is an overhead for passing the results from the subgraph to the main graph.
The above logic is already used by the TensorFlow Lite GPU backend described in the next section.

\subsection{TensorFlow Lite GPU Delegate}

While many different NPUs were already released by all major players, they are still very fragmented due to a missing common interface or API. While NNAPI was designed to tackle this problem, it suffers from its own design flaws that slow down NNAPI adoption and usage growth:

\vspace{-0.8mm}
\begin{itemize}
\setlength\itemsep{-0.2mm}
\item \textBF{Long update cycle:}\, NNAPI update is still bundled with the OS update. Thus, it may take up to a year to get new drivers.
\item \textBF{Custom operations support:}\, When a model has an op that is not yet supported by NNAPI, it is nearly impossible to run it with NNAPI. In the worst case, two parts of a graph are accelerated through NNAPI, while a single op implemented out of the context is computed by the CPU, which ruins the performance.
\end{itemize}
\vspace{-0.8mm}

There is another attempt by the Vulkan ML group to introduce common programming language to be implemented by vendors. The language resembles a model graph representation similar to one found in the TensorFlow or ONNX libraries.
%Vendors could use GPUs and / or NPUs in their implementations.
The proposal is still in its early stage and, if accepted, will take a few years to reach consumer devices.
%Besides that, Google introduced ML Intermediate Representation (MLIR) that has similar to the LLVM compiler goal but applicable to NN/ML world: let framework developers to map NN layers to lower representations that in turn could be translated into a binary code executable on CPU, DSP, NPU, GPU. MLIR-based implementations are still yet to be released to prove the point.

Besides the above issues, there also exists a huge fragmentation of mobile hardware platforms. For instance, the most popular 30 SoC designs are now representing only 51\% of the market share, while 225 SoCs are still covering just 95\% of the market with a long tail of a few thousand designs. The majority of these SoCs will never get NNAPI drivers, though one should mention that around 23\% of them have GPUs at least 2 times more performant than the corresponding CPUs, and thus they can be used for accelerating ML inference. This number is significantly bigger than the current market share of chipsets with NPUs or valid NNAPI drivers. To use the GPU acceleration on such platforms, TensorFlow GPU delegate was introduced.

The inference phase of the GPU delegate consists of the following steps. The input tensors are first reshaped to the PHWC4 format if their tensor shape has channel size not equal to 4. For each operator, shader programs are linked by binding resources such the operator's input / output tensors, weights, etc. and dispatched, i.e.~inserted into the command queue. The GPU driver then takes care of scheduling and executing all shader programs in the queue, and makes the result available to the CPU by the CPU / GPU synchronization. In the GPU inference engine, operators exist in the form of shader programs. The shader programs eventually get compiled and inserted into the command queue and the GPU executes programs from this queue without synchronization with the CPU.
After the source code for each program is generated, each shader gets compiled. This compilation step can take awhile, from several milliseconds to seconds. Typically, app developers can hide this latency while loading the model or starting the app for the first time. Once all shader programs are compiled, the GPU backend is ready for inference. A much more detailed description of the TFlite GPU delegate can be found in~\cite{lee2019device}.

\begin{figure*}[t!]
\centering
\resizebox{1.0\linewidth}{!}
{
\includegraphics[width=1.0\linewidth]{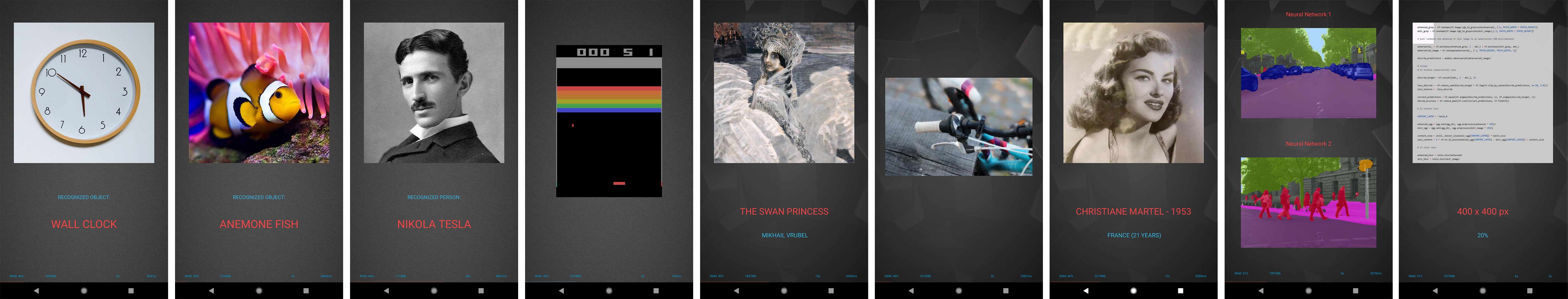}
}
\vspace{0mm}
\caption{\small{Sample result visualizations displayed to the user in deep learning tests.}}
\vspace{-2.2mm}
\label{fig:tests}
\end{figure*}

\subsection{Floating-point vs. Quantized Inference}

One of the most controversial topics related to the deployment of deep learning models on smartphones is the suitability of floating-point and quantized models for mobile devices. There has been a lot of confusion with these two types in the mobile industry, including a number of incorrect statements and invalid comparisons. We therefore decided to devote a separate section to them and describe and compare their benefits and disadvantages. We divided the discussion into three sections: the first two are describing each inference type separately, while the last one compares them directly and makes suggestions regarding their application.

\subsubsection{Floating-point Inference}

\hspace{3.2mm} \textBF{Advantages:}\,
The model is running on mobile devices in the same format as it was originally trained on the server or desktop with standard machine learning libraries. No special conversion, changes or re-training is needed; thus one gets the same accuracy and performance as on the desktop or server environment.

\smallskip

\textBF{Disadvantages:}\,
Many recent state-of-the-art deep learning models, especially those that are working with high-resolution image transformations, require more than 6-8 gigabytes of RAM and enormous computational resources for data processing that are not available even in the latest high-end smartphones. Thus, running such models in their original format is infeasible, and they should be first modified to meet the hardware resources available on mobile devices.

\subsubsection{Quantized Inference}

\hspace{3.2mm} \textBF{Advantages:}\,
The model is first converted from a 16-bit floating point type to int-8 format. This reduces its size and RAM consumption by a factor of 4 and potentially speeds up its execution by 2-3 times. Since integer computations consume less energy on many platforms, this also makes the inference more power efficient, which is critical in the case of smartphones and other portable electronics.

\smallskip

\textBF{Disadvantages:}\,
Reducing the bit-width of the network weights (from 16 to 8 bits) leads to accuracy loss: in some cases, the converted model might show only a small performance degradation, while for some other tasks the resulting accuracy will be close to zero. Although a number of research papers dealing with network quantization were presented by Qualcomm~\cite{louizos2018relaxed,nagel2019data} and Google~\cite{jacob2018quantization,krishnamoorthi2018quantizing}, all showing decent accuracy results for many image classification models, there is no general recipe for quantizing arbitrary deep learning architectures. Thus, quantization is still more of a research topic, without working solutions for many AI-related tasks (\eg image-to-image mapping or various NLP problems). Besides that, many quantization approaches require the model to be retrained from scratch, preventing the developers from using available pre-trained models provided together with all major research papers.

\subsubsection{Comparison}

As one can see, there is always a trade-off between using one model type or another: floating-point models will always show better accuracy (since they can be simply initialized with the weights of the quantized model and further trained for higher accuracy), while integer models yield faster inference. The progress here comes from both sides: AI accelerators for floating-point models are becoming faster and are reducing the difference between the speed of INT-8 and FP16 inference, while the accuracy of various network quantization approaches is also rising rapidly. Thus, the applicability of each approach will depend on the particular task and the corresponding hardware / energy consumption limitations: for complex models and high-performance devices float models are preferable (due to the convenience of deployment and better accuracy), while quantized inference is clearly beneficial in the case of low-power and low-RAM devices and quantization-friendly models that can be converted from the original float format to INT-8 with a minimal performance degradation.

When comparing float and quantized inference, one good analogy would be the use of FullHD vs.~4K videos on mobile devices. All other things being equal, the latter always have better quality due to their higher resolution, but also demand considerably more disc space or internet bandwidth and hardware resources for decoding them. Besides that, on some screens the difference between 1080P and 4K might not be visible. But this does not mean that one of the two resolutions should be discarded altogether. Rather, the most suitable solution should be used in each case.

Last but not least, one should definitely avoid comparing the performance of two different devices by running floating-point models on one and quantized models on the other. As they have different properties and show different accuracy results, the obtained numbers will make no sense (same as measuring the FPS in a video game running on two devices with different resolutions). This, however, does not refer to the situation when this is done to demonstrate the comparative performance of two inference types, if accompanied by the corresponding accuracy results.

\vspace{-2mm}
\section{AI Benchmark 3.0}
\label{sec:ai_benchmark_3}

The AI Benchmark application was first released in May 2018, with the goal of measuring the AI performance of various mobile devices. The first version (1.0.0) included a number of typical AI tasks and deep learning architectures, and was measuring the execution time and memory consumption of the corresponding AI models. In total, 12 public versions of the AI Benchmark application were released since the beginning of the project. The second generation (2.0.0) was described in detail in the preceding paper~\cite{ignatov2018ai}. Below we briefly summarize the key changes introduced in the subsequent benchmark releases:

\smallskip

\noindent~--\, \textBF{2.1.0}\, \textit{(release date: 13.10.2018)}~--- this version brought a number of major changes to AI Benchmark. The total number of tests was increased from 9 to 11. In test 1, MobileNet-V1 was changed to MobileNet-V2 running in three subtests with different inference types: float model on CPU, float model with NNAPI and quantized model with NNAPI. Inception-ResNet-V1 and VGG-19 models from tests 3 and 5, respectively, were quantized and executed with NNAPI. In test 7, ICNet model was running in parallel in two separate threads on CPU. A more stable and reliable category-based scoring system was introduced. Required Android 4.1 and above.

\smallskip

\noindent~--\, \textBF{2.1.1}\, \textit{(release date: 15.11.2018)}~--- normalization coefficients used in the scoring system were updated to be based on the best results from the actual SoCs generation (Snapdragon 845, Kirin 970, Helio P60 and Exynos 9810). This version also introduced several bug fixes and an updated ranking table. Required Android 4.1 and above.

\smallskip

\noindent~--\, \noindent\textBF{2.1.2}\, \textit{(release date: 08.01.2019)}~--- contained a bug fix for the last memory test (on some devices, it was terminated before the actual RAM exhaustion).

\smallskip

\noindent~--\, \noindent\textBF{3.0.0}\, \textit{(release date: 27.03.2019)}~--- the third version of AI Benchmark with a new modular-based architecture and a number of major updates. The number of tests was increased from 11 to 21. Introduced accuracy checks, new tasks and networks, PRO mode and updated scoring system that are described further in this section.

\smallskip

\noindent~--\, \noindent\textBF{3.0.1}\, \textit{(release date: 21.05.2019)} and \textBF{3.0.2}\, \textit{(release date: 13.06.2019)}~--- fixed several bugs and introduced new features in the PRO mode.

\smallskip

Since a detailed technical description of AI Benchmark 2.0 was provided in~\cite{ignatov2018ai}, we here mainly focus on the updates and changes introduced by the latest release.

\begin{table*}[t]
\centering
\resizebox{2.0\columnwidth}{!}
{
\begin{tabular}{l|cccc|cccccc}
Test & 1 & 2 & 3 & 4 & 5 & 6 & 7 & 8 & 9 & 10 \\[0.1cm]
\hline\\
Task \, & \, Classification \, & \, Classification \, & \, Face Recognition \, & \, Playing Atari \, & \, Deblurring \, & \, Super-Resolution \, & \, Super-Resolution \, & \, Bokeh Simulation \, & \, Segmentation \, & \, Enhancement \, \\
Architecure \, & MobileNet-V2 & Inception-V3 & Inc-ResNet-V1 & LSTM RNN & SRCNN & VGG-19 & SRGAN (ResNet-16) & U-Net & ICNet & DPED (ResNet-4) \\
Resolution, px \, & \, 224$\times$224 \, & \, 346$\times$346 \, & \, 512$\times$512 \, & \, 84$\times$84 \, & \, 384$\times$384 \, & \, 256$\times$256 \, & \, 512$\times$512 \, & \, 128$\times$128 \, & \, 768$\times$1152 \, & \, 128$\times$192 \\
Parameters \, & \, 3.5M \, & \, 27.1M \, & \, 22.8M \, & \, 3.4M \, & \, 69K \, & \, 665K \, & \, 1.5M \, & \, 6.6M \, & \, 6.7M \, & \, 400K \\
Size (float), MB \, & \, 14 \, & \, 95 \, & \, 91 \, & \, 14 \, & \, 0.3 \, & \, 2.7 \, & \, 6.1 \, & \, 27 \, & \, 27 \, & \, 1.6 \\
NNAPI support \, & \, yes \, & \, yes \, & \, yes \, & \, yes (1.2+) \, & \, yes \, & \, yes \, & \, yes (1.2+) \, & \, yes (1.2+) & \, yes \, & \, yes \\
CPU-Float \, & \, yes \, & \, yes \, & \, no \, & \, yes \, & \, no \, & \, no \, & \, yes \, & \, yes & \, no \, & \, no \\
CPU-Quant \, & \, no \, & \, no \, & \, yes \, & \, no \, & \, no \, & \, no \, & \, yes \, & \, no & \, no \, & \, no \\
NNAPI-Float \, & \, yes \, & \, yes \, & \, yes \, & \, no \, & \, yes \, & \, yes \, & \, no \, & \, no & \, yes \, & \, yes \\
NNAPI-Quant \, & \, yes \, & \, yes \, & \, yes \, & \, no \, & \, yes \, & \, yes \, & \, no \, & \, no & \, no \, & \, no \\
\end{tabular}
}
\vspace{2.6mm}
\caption{Summary of deep learning models used in the AI Benchmark.}
\vspace{2.6mm}
\label{networks-summary}
\end{table*}

\vspace{-0.8mm}
\subsection{Deep Learning Tests}

The actual benchmark version (3.0.2) consists of 11 test sections and 21 tests. The networks running in these tests represent the most popular and commonly used deep learning architectures that can be currently deployed on smartphones. The description of test configs is provided below.

\bigskip

\noindent \textBF{Test Section 1:}\,\, Image Classification

\smallskip
\small

Model:\, MobileNet-V2~\cite{sandler2018mobilenetv2},

Inference modes:\, CPU (FP16/32) and NNAPI (INT8 + FP16)

Image resolution:\, 224$\times$224 px, \, Test time limit:\, 20 seconds

\normalsize
\smallskip

\noindent \textBF{Test Section 2:}\,\, Image Classification

\smallskip
\small

Model:\, Inception-V3~\cite{szegedy2016rethinking}

Inference modes:\, CPU (FP16/32) and NNAPI (INT8 + FP16)

Image resolution:\, 346$\times$346 px, \, Test time limit:\, 30 seconds

\normalsize
\smallskip

\noindent \textBF{Test Section 3:}\,\, Face Recognition

\smallskip
\small

Model:\, Inception-ResNet-V1~\cite{szegedy2017inception}

Inference modes:\, CPU (INT8) and NNAPI (INT8 + FP16)

Image resolution:\, 512$\times$512 px, \, Test time limit:\, 30 seconds

\normalsize
\smallskip

\noindent \textBF{Test Section 4:}\,\, Playing Atari

\smallskip
\small

Model:\, LSTM~\cite{hochreiter1997long}

Inference modes:\, CPU (FP16/32)

Image resolution:\, 84$\times$84 px, \, Test time limit:\, 20 seconds

\normalsize
\smallskip

\noindent \textBF{Test Section 5:}\,\, Image Deblurring

\smallskip
\small

Model:\, SRCNN 9-5-5~\cite{dong2016image}

Inference modes:\, NNAPI (INT8 + FP16)

Image resolution:\, 384$\times$384 px, \, Test time limit:\, 30 seconds

\normalsize
\smallskip

\noindent \textBF{Test Section 6:}\,\, Image Super-Resolution

\smallskip
\small

Model:\, VGG-19 (VDSR)~\cite{kim2016accurate}

Inference modes:\, NNAPI (INT8 + FP16)

Image resolution:\, 256$\times$256 px, \, Test time limit:\, 30 seconds

\normalsize
\smallskip

\noindent \textBF{Test Section 7:}\,\, Image Super-Resolution

\smallskip
\small

Model:\, SRGAN~\cite{ledig2017photo}

Inference modes:\, CPU (INT8 + FP16/32)

Image resolution:\, 512$\times$512 px, \, Test time limit:\, 40 seconds

\normalsize
\smallskip

\noindent \textBF{Test Section 8:}\,\, Bokeh Simulation

\smallskip
\small

Model:\, U-Net~\cite{ronneberger2015u}

Inference modes:\, CPU (FP16/32)

Image resolution:\, 128$\times$128 px, \, Test time limit:\, 20 seconds

\normalsize
\smallskip

\noindent \textBF{Test Section 9:}\,\, Image Segmentation

\smallskip
\small

Model:\, ICNet~\cite{zhao2017icnet}

Inference modes:\, NNAPI (2 $\times$ FP32 models in parallel)

Image resolution:\, 768$\times$1152 px, \, Test time limit:\, 20 seconds

\normalsize
\smallskip

\noindent \textBF{Test Section 10:}\,\, Image Enhancement

\smallskip
\small

Model:\, DPED-ResNet~\cite{ignatov2017dslr,ignatov2018wespe}

Inference modes:\, NNAPI (FP16 + FP32)

Image resolution:\, 128$\times$192 px, \, Test time limit:\, 20 seconds

\normalsize
\smallskip

\noindent \textBF{Test Section 11:}\,\, Memory Test

\smallskip
\small

Model:\, SRCNN 9-5-5~\cite{dong2016image}

Inference modes:\, NNAPI (FP16)

Image resolution:\, from 200$\times$200 px to 2000$\times$2000 px

\normalsize
\smallskip

\begin{figure}[t!]
\centering
\resizebox{0.8\linewidth}{!}
{
\includegraphics[width=1.0\linewidth]{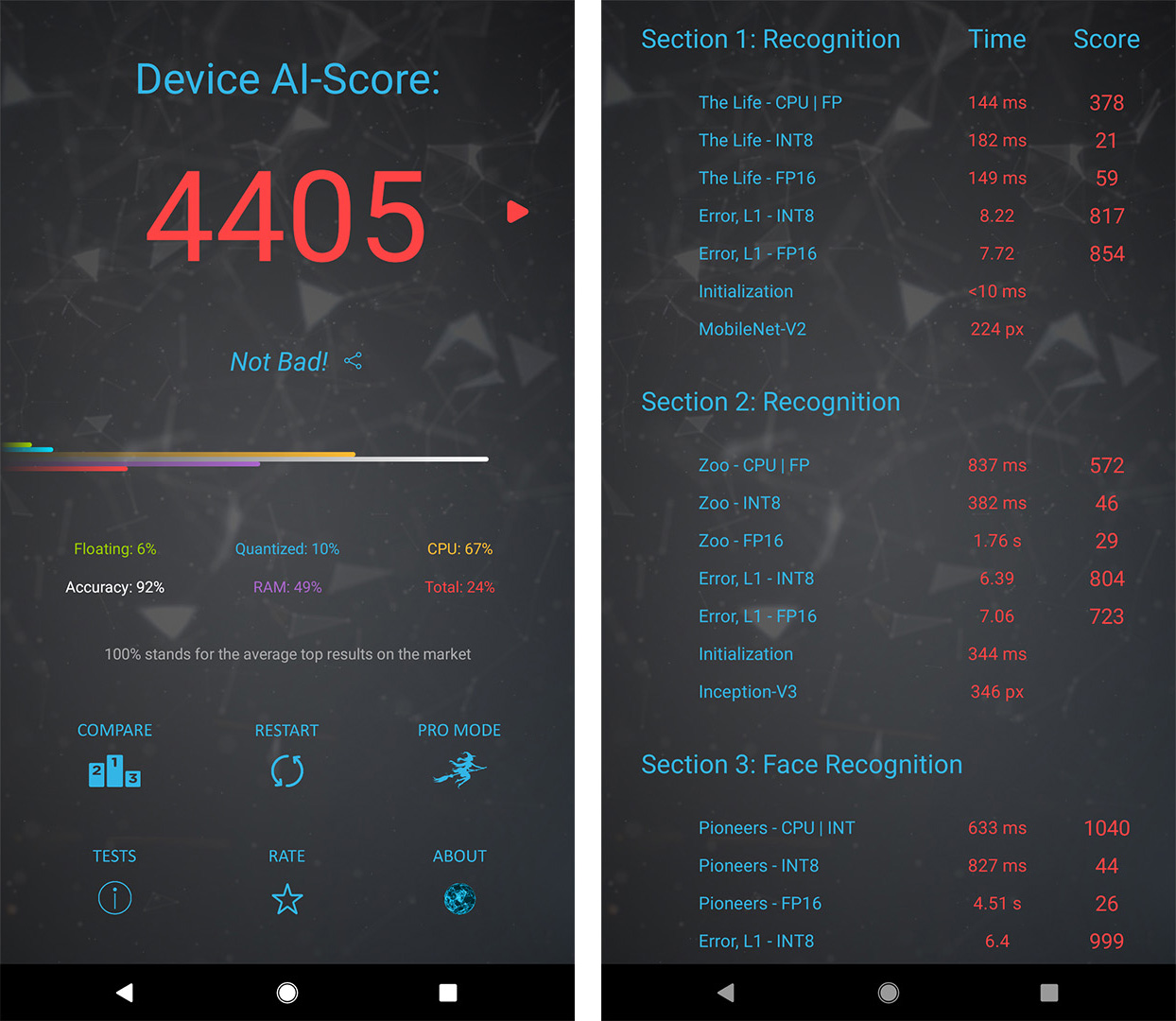}
}
\vspace{2.6mm}
\caption{\small{Benchmark results displayed after the end of the tests.}}
\vspace{-1.8mm}
\label{fig:results-vis}
\end{figure}

Table~\ref{networks-summary} summarizes the details of all the deep learning architectures included in the benchmark. When more than one inference mode is used, each image is processed sequentially with all the corresponding modes. In the last memory test, images are processed until the Out-Of-Memory-Error is thrown or all resolutions are processed successfully. In the image segmentation test (Section 9), two TFLite ICNet models are initialized in two separate threads and process images in parallel (asynchronously) in these two threads. The running time for each test is computed as an average over the set of images processed within the specified time. When more than two images are processed, the first two results are discarded to avoid taking into account initialization time (estimated separately), and the average over the rest results is calculated. If less than three images are processed (which happens only on low-end devices), the last inference time is used. The benchmark's visualization of network outputs is shown in Fig.~\ref{fig:tests}.

Starting from version 3.0.0, AI Benchmark is checking the accuracy of the outputs for float and quantized models running with acceleration (NNAPI) in Test Sections 1, 2, 3, 5 and 6. For each corresponding test, the $L_1$ loss is computed between the target and actual outputs produced by the deep learning models. The accuracy is estimated separately for both float and quantized models.

\subsection{Scoring System}

AI Benchmark is measuring the performance of several test categories, including int-8, float-16, float-32, parallel, CPU (int-8 and float-16/32), memory tests, and tests measuring model initialization time. The scoring system used in versions 3.0.0~-- 3.0.2 is identical. The contribution of the test categories is as follows:

\vspace{-2mm}
\begin{itemize}
\setlength\itemsep{-0.2mm}
\item 48\% - float-16 tests;
\item 24\% - int-8 tests;
\item 12\% - CPU, float-16/32 tests;
\item 6\% - CPU, int-8 tests;
\item 4\% - float-32 tests;
\item 3\% - parallel execution of the models;
\item 2\% - initialization time, float models;
\item 1\% - initialization time, quantized models;
\end{itemize}
\vspace{-2mm}

The scores of each category are computed as a geometric mean of the test results belonging to this category. The computed L1 error is used to penalize the runtime of the corresponding networks running with NNAPI (an exponential penalty with exponent 1.5 is applied). The result of the memory test introduces a multiplicative contribution to the final score, displayed at the end of the tests (Fig.~\ref{fig:results-vis}). The normalization coefficients for each test are computed based on the best results of the current SoC generation (Snapdragon 855, Kirin 980, Exynos 9820 and Helio P90).

\smallskip

\subsection{PRO Mode}

\begin{figure}[t!]
\centering
\resizebox{1.0\linewidth}{!}
{
\includegraphics[width=1.0\linewidth]{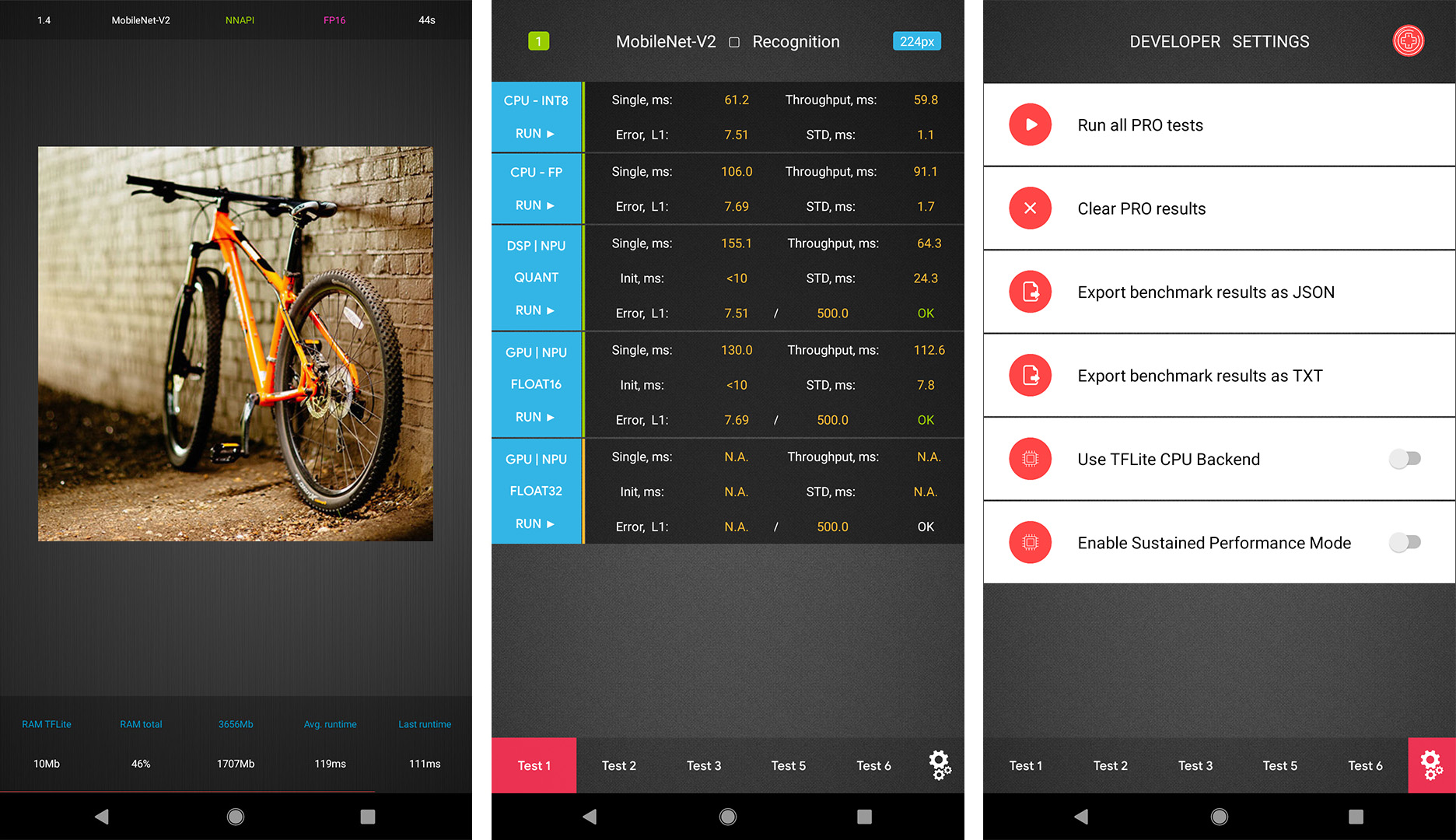}
}
\vspace{-1.2mm}
\caption{\small{Tests, results and options displayed in the PRO Mode.}}
\vspace{-1.8mm}
\label{fig:results-pro}
\end{figure}

The PRO Mode (Fig.~\ref{fig:results-pro}) was introduced in AI Benchmark 3.0.0 to provide developers and experienced users with the ability to get more detailed and accurate results for tests running with acceleration, and to compare the results of CPU- and NNAPI-based execution for all inference types. It is available only for tasks where both the float and quantized models are compatible with NNAPI (Test Sections 1, 2, 3, 5, 6). In this mode, one can run each of the five inference types (CPU-float, CPU-quantized, float-16-NNAPI, float-32-NNAPI and int-8-NNAPI) to get the following results:

\vspace{-0.8mm}
\begin{itemize}
\setlength\itemsep{-0.2mm}
\item Average inference time for a single-image inference;
\item Average inference time for a throughput inference;
\item Standard deviation of the results;
\item The accuracy of the produced outputs ($L_1$ error);
\item Model's initialization time.
\end{itemize}
\vspace{-0.8mm}

Some additional options were added to the PRO Mode in version 3.0.1 that are available under the ``Settings'' tab:

\begin{enumerate}
\setlength\itemsep{-0.2mm}
\item All PRO Mode tests can be run in automatic mode;
\item Benchmark results can be exported to a JSON / TXT file stored in the device's internal memory;
\item TensorFlow Lite CPU backend can be enabled in all tests for debugging purposes;
\item Sustained performance mode can be used in all tests.
\end{enumerate}

\vspace{-2mm}
\subsection{AI Benchmark for Desktops}

Besides the Android version, a separate open source AI Benchmark build for desktops~\footnote{\url{http://ai-benchmark.com/alpha}} was released in June 2019. It is targeted at evaluating AI performance of the common hardware platforms, including CPUs, GPUs and TPUs, and measures the inference and training speed for several key deep learning models. The benchmark is relying on the TensorFlow~\cite{abadi2016tensorflow} machine learning library and is distributed as a Python pip package~\footnote{\url{https://pypi.org/project/ai-benchmark}} that can be installed on any system running Windows, Linux or macOS. The current release 0.1.1 consists of 42 tests and 19 sections provided below:

\begin{table*}[t!]
\centering
\vspace{2mm}
\resizebox{1.9\columnwidth}{!}
{
\begin{tabular}{l|c|cccccc|c}
SoC \, & \, AI Accelerator \, & \, MobileNet \, & \, Inception \, & \, Inc-ResNet \, & \, SRCNN, \, & \, VGG-19, \, & \, DPED, \, & \,\, Relative \\
&  & v2, ms & v3, ms & v1, ms & ms & ms & ms & Perf. \\ [0.1cm]
\hline\\
HiSilicon Kirin 990 & NPU (Da Vinci family) & \textBF{6} & \textBF{18} & \textBF{37} & \textBF{36} & \textBF{42} & \textBF{19} & 100\% \\
HiSilicon Kirin 810 & NPU (Da Vinci family) & 10 & 34 & 82 & 72 & 122 & 42 & 47\% \\
Unisoc Tiger T710 & NPU & 13 & 35 & 80 & 76 & 135 & 43 & 43\% \\
Snapdragon 855 Plus & GPU (Adreno 640) & 15 & 56 & 142 & 66 & 182 & 67 & 32\% \\
HiSilicon Kirin 980 & NPU (Cambricon family) & 25 & 58 & 117 & 100 & 163 & 71 & 28\% \\
Exynos 9825 Octa & GPU (Mali-G76 MP12, S.LSI OpenCL) & 17 & 65 & 151 & 124 & 158 & 67 & 28\% \\
MediaTek Helio P90 & APU 2.0 & 8.3 & 101 & 263 & 75 & 309 & 66 & 26\% \\
Exynos 9820 Octa & GPU (Mali-G76 MP12, S.LSI OpenCL) & 19 & 69 & 162 & 137 & 170 & 74 & 26\% \\
Snapdragon 855 & GPU (Adreno 640) & 24 & 70 & 170 & 87 & 211 & 82 & 25\% \\
Snapdragon 845 & GPU (Adreno 630) & 27 & 80 & 205 & 98 & 263 & 94 & 21\% \\
HiSilicon Kirin 970 & NPU (Cambricon family) & 43 & 69 & 1514 & 141 & 235 & 83 & 14\% \\
Snapdragon 730 & GPU (Adreno 618) & 31 & 150 & 391 & 185 & 553 & 175 & 12\% \\
MediaTek Helio G90T & GPU (Mali-G76 MP4) & 37 & 223 & 584 & 459 & 977 & 665 & 6\% \\
Exynos 9820 Octa & GPU (Mali-G76 MP12, Arm NN OpenCL) & 40 & 186 & 442 & 889 & 837 & 836 & 6\% \\
Snapdragon 675 & GPU (Adreno 612) & 39 & 312 & 887 & 523 & 1238 & 347 & 6\% \\
Exynos 9810 Octa & GPU (Mali-G72 MP18) & 72 & 209 & 488 & 1574 & 843 & 787 & 4\% \\
Exynos 8895 Octa & GPU (Mali-G71 MP20) & 63 & 216 & 497 & 1785 & 969 & 909 & 4\% \\
MediaTek Helio P70 & GPU (Mali-G72 MP3) & 66 & 374 & 932 & 1096 & 865 & 764 & 4\% \\
MediaTek Helio P65 & GPU (Mali-G52 MP2) & 51 & 340 & 930 & 752 & 1675 & 926 & 4\% \\
Snapdragon 665 & GPU (Adreno 610) & 50 & 483 & 1292 & 678 & 2174 & 553 & 4\% \\
MediaTek Helio P60 & GPU (Mali-G72 MP3) & 68 & 353 & 948 & 1896 & 889 & 1439 & 3\% \\
Exynos 9609 & GPU (Mali-G72 MP3) & 61 & 444 & 1230 & 1661 & 1448 & 731 & 3\% \\
Exynos 9610 & GPU (Mali-G72 MP3) & 77 & 459 & 1244 & 1651 & 1461 & 773 & 3\% \\
Exynos 8890 Octa & GPU (Mali-T880 MP12) & 98 & 447 & 1012 & 2592 & 1062 & 855 & 3\% \\
\hline
Snapdragon 835 & None & 181 & 786 & 1515 & 1722 & 3754 & 1317 & 1\% \\
\hline
GeForce GTX 1080 Ti & CUDA (3584 cores, 1.58 - 1.60 GHz) & 1.5 & 4.5 & 9.5 & 4.7 & 10 & 4.6 & 449\% \\
GeForce GTX 950 & CUDA (768 cores, 1.02 - 1.19 GHz) & 3.9 & 15 & 38 & 23 & 47 & 20 & 115\% \\
Nvidia Tesla K40c & CUDA (2880 cores, 0.75 - 0.88 GHz) & 3.7 & 16 & 38 & 22 & 60 & 20 & 111\% \\
Quadro M2000M & CUDA (640 cores, 1.04 - 1.20 GHz) & 5 & 22 & 54 & 33 & 84 & 30 & 78\% \\
GeForce GT 1030 & CUDA (384 cores, 1.23 - 1.47 GHz) & 9.3 & 31 & 81 & 44 & 97 & 47 & 53\% \\
GeForce GT 740 & CUDA (384 cores, 0.993 GHz) & 12 & 89 & 254 & 238 & 673 & 269 & 14\% \\
GeForce GT 710 & CUDA (192 cores, 0.954 GHz) & 33 & 159 & 395 & 240 & 779 & 249 & 10\% \\
\hline
Intel Core i7-9700K & 8/8 @ 3.6 - 4.9 GHz, Intel MKL & 4.8 & 23 & 72 & 49 & 133 & 72 & 55\% \\
Intel Core i7-7700K & 4/8 @ 4.2 - 4.5 GHz, Intel MKL & 7.4 & 42 & 121 & 75 & 229 & 100 & 34\% \\
Intel Core i7-4790K & 4/8 @ 4.0 - 4.4 GHz, Intel MKL & 8.3 & 45 & 133 & 91 & 267 & 124 & 30\% \\
Intel Core i7-3770K & 4/8 @ 3.5 - 3.9 GHz, Intel MKL & 12 & 125 & 345 & 209 & 729 & 242 & 13\% \\
Intel Core i7-2600K & 4/8 @ 3.4 - 3.8 GHz, Intel MKL & 14 & 143 & 391 & 234 & 816 & 290 & 11\% \\
Intel Core i7-950 & 4/8 @ 3.1 - 3.3 GHz, Intel MKL & 36 & 287 & 728 & 448 & 1219 & 515 & 6\% \\
\end{tabular}
}
\vspace{1.6mm}
\caption{Inference time (per one image) for \textbf{floating-point networks} obtained on mobile SoCs providing hardware acceleration for fp-16 models. The results of the Snapdragon 835, Intel CPUs and Nvidia GPUs are provided for reference. Acceleration on Intel CPUs was achieved using the Intel MKL-DNN library~\cite{MKLDNN2019}, on Nvidia GPUs~-- with CUDA~\cite{CUDA2019} and cuDNN~\cite{chetlur2014cudnn}. The results on Intel and Nvidia hardware were obtained using the standard TensorFlow library~\cite{abadi2016tensorflow} running floating-point models with a batch size of 10. A full list is available at:\, \small{\url{http://ai-benchmark.com/ranking_processors}}}
\label{ranking-phones-float}
\vspace{0mm}
\end{table*}

\begin{table*}[t!]
\centering
\resizebox{1.9\columnwidth}{!}
{
\begin{tabular}{l|c|ccccc|c}
SoC \, & \, AI Accelerator \, & \, MobileNet \, & \, Inception \, & \, Inc-ResNet \, & \, SRCNN, \, & \, VGG-19, \, & \,\, Relative \\
&  & v2, ms & v3, ms & v1, ms & ms & ms & Perf. \\ [0.1cm]
\hline\\
Snapdragon 855 Plus & DSP (Hexagon 690) & 4.9 & \textBF{16} & 40 & 24 & 45 & 100\% \\
Unisoc Tiger T710 & NPU & 5 & 17 & 38 & \textBF{20} & 53 & 99\% \\
HiSilicon Kirin 990 & NPU (Da Vinci family) & 6.5 & 20 & \textBF{37} & 38 & \textBF{39} & 86\% \\
Snapdragon 855 & DSP (Hexagon 690) & 8.2 & 18 & 46 & 30 & 48 & 80\% \\
MediaTek Helio P90 & APU 2.0 & \textBF{4} & 23 & 38 & 22 & 147 & 78\% \\
Snapdragon 675 & DSP (Hexagon 685) & 10 & 34 & 73 & 53 & 103 & 47\% \\
Snapdragon 730 & DSP (Hexagon 688) & 13 & 47 & 90 & 69 & 111 & 38\% \\
%Snapdragon 675 & DSP (Hexagon 685) & 13 & 41 & 90 & 67 & 134 & 37\% \\
Snapdragon 670 & DSP (Hexagon 685) & 12 & 48 & 97 & 153 & 116 & 32\% \\
Snapdragon 665 & DSP (Hexagon 686) & 13 & 52 & 118 & 94 & 192 & 29\% \\
Snapdragon 845 & DSP (Hexagon 685) & 11 & 45 & 91 & 71 & 608 & 28\% \\
Exynos 9825 Octa & GPU (Mali-G76 MP12, S.LSI OpenCL) & 19 & 63 & 128 & 75 & 199 & 27\% \\
Snapdragon 710 & DSP (Hexagon 685) & 12 & 48 & 95 & 70 & 607 & 27\% \\
MediaTek Helio G90T & APU 1.0 & 15 & 64 & 139 & 107 & 308 & 23\% \\
Exynos 9820 Octa & GPU (Mali-G76 MP12, S.LSI OpenCL) & 21 & 73 & 199 & 87 & 262 & 21\% \\
HiSilicon Kirin 810 & NPU (Da Vinci family) & 25 & 98 & 160 & 116 & 172 & 21\% \\
MediaTek Helio P70 & APU 1.0 & 26 & 89 & 181 & 163 & 474 & 15\% \\
MediaTek Helio P60 & APU 1.0 & 27 & 89 & 181 & 164 & 475 & 15\% \\
Exynos 9820 Octa & GPU (Mali-G76 MP12, Arm NN OpenCL) & 27 & 96 & 201 & 407 & 446 & 12\% \\
Exynos 8895 Octa & GPU (Mali-G71 MP20) & 44 & 118 & 228 & 416 & 596 & 10\% \\
Exynos 9810 Octa & GPU (Mali-G72 MP18) & 45 & 166 & 360 & 539 & 852 & 7\% \\
MediaTek Helio P65 & GPU (Mali-G52 MP2) & 43 & 228 & 492 & 591 & 1167 & 6\% \\
\hline
Snapdragon 835 & None & 136 & 384 & 801 & 563 & 1525 & 3\% \\
\hline
Exynos 9609 & GPU (Mali-G72 MP3) & 50 & 383 & 937 & 1027 & 2325 & 3\% \\
Exynos 9610 & GPU (Mali-G72 MP3) & 52 & 380 & 927 & 1024 & 2322 & 3\% \\
Exynos 8890 Octa & GPU (Mali-T880 MP12) & 70 & 378 & 866 & 1200 & 2016 & 3\% \\
\end{tabular}
}
\vspace{1.6mm}
\caption{Inference time for \textbf{quantized networks} obtained on mobile SoCs providing hardware acceleration for int-8 models. The results of the Snapdragon 835 are provided for reference. A full list is available at:\, \small{\url{http://ai-benchmark.com/ranking_processors}}}

\label{ranking-phones-quant}
\vspace{0.2mm}
\end{table*}

\vspace{-0.6mm}
\begin{enumerate}
\setlength\itemsep{-0.2mm}
\item MobileNet-V2~\cite{sandler2018mobilenetv2}\, [classification]
\item Inception-V3~\cite{szegedy2016rethinking}\, [classification]
\item Inception-V4~\cite{szegedy2017inception}\, [classification]
\item Inception-ResNet-V2~\cite{he2016identity}\, [classification]
\item ResNet-V2-50~\cite{he2016identity}\, [classification]
\item ResNet-V2-152~\cite{he2016identity}\, [classification]
\item VGG-16~\cite{simonyan2014very}\, [classification]
\item SRCNN 9-5-5~\cite{dong2016image}\, [image-to-image mapping]
\item VGG-19~\cite{kim2016accurate}\, [image-to-image mapping]
\item ResNet-SRGAN~\cite{ledig2017photo}\, [image-to-image mapping]
\item ResNet-DPED~\cite{ignatov2017dslr,ignatov2018wespe}\, [image-to-image mapping]
\item U-Net~\cite{ronneberger2015u}\, [image-to-image mapping]
\item Nvidia-SPADE~\cite{park2019semantic}\, [image-to-image mapping]
\item ICNet~\cite{zhao2017icnet}\, [image segmentation]
\item PSPNet~\cite{zhao2017pyramid}\, [image segmentation]
\item DeepLab~\cite{papandreou2015weakly}\, [image segmentation]
\item Pixel-RNN~\cite{oord2016pixel}\, [inpainting]
\item LSTM~\cite{hochreiter1997long}\, [sentence sentiment analysis]
\item GNMT~\cite{wu2016google} [text\, translation]
\end{enumerate}
\vspace{-1.2mm}

The results obtained with this benchmark version are available on the project webpage~\footnote{\url{http://ai-benchmark.com/ranking_deeplearning}}. Upcoming releases will provide a unified ranking system that allows for a direct comparison of results on mobile devices (obtained with Android AI Benchmark) with those on desktops. The current constraints and particularities of mobile inference do not allow us to merge these two AI Benchmark versions right now, however, they will be gradually consolidated into a single AI Benchmark Suite with a global ranking table. The numbers for desktop GPUs and CPUs shown in the next section were obtained with a modified version of the desktop AI Benchmark build.

\section{Benchmark Results}
\label{sec:benchmark_results}

As the performance of mobile AI accelerators has grown significantly in the past year, we decided to add desktop CPUs and GPUs used for training / running deep learning models to the comparison as well. This will help us to understand how far mobile AI silicon has progressed thus far. It also will help developers to estimate the relation between the runtime of their models on smartphones and desktops. In this section, we present quantitative benchmark results obtained from over 20,000 mobile devices tested in the wild (including a number of prototypes) and discuss in detail the performance of all available mobile chipsets providing hardware acceleration for floating-point or quantized models. The results for floating-point and quantized inference obtained on mobile SoCs are presented in tables~\ref{ranking-phones-float} and~\ref{ranking-phones-quant}, respectively. The detailed performance results for smartphones are shown in table~\ref{ranking-phones-global}.

\begin{table*}[t!]
\centering
\resizebox{2.0\columnwidth}{!}
{
\begin{tabular}{l|c|cccccccccccccccccccccccccccccc|c|c}
Phone Model & SoC & 1c-f, & 1q, & 1q, & 1f, & 1f, & 2c-f, & 2q, & 2q, & 2f, & 2f, & 3c-f, & 3q, & 3q, & 3f, & 3f, & 4c-f, & 5q, & 5q, & 5f, & 5f, & 6q, & 6q, & 6f, & 6f, & 7c-q, & 7c-f, & 8c-fq, & 9f-p, & 10f, & 10f32, & 11-m, & AI-Score \\
& & ms & ms & error & ms & error & ms & ms & error & ms & error & ms & ms & error & ms & error & ms & ms & error & ms & error & ms & error & ms & error & ms & ms & ms & ms & ms & ms & px \\
\hline\\
Huawei Mate 30 Pro 5G & HiSilicon Kirin 990 & 38 & 6.5 & 7.7 & 6 & 7.78 & 538 & 20 & 5.36 & 18 & 5.59 & 961 & 37 & 13.9 & 37 & 7.33 & 86 & 38 & 4.36 & 36 & 3.23 & 39 & 3.82 & 42 & 3.06 & 717 & 1627 & 1196 & 234 & 19 & 218 & 2000 & 76206 \\
Honor 9X Pro & HiSilicon Kirin 810 & 38 & 25 & 6.99 & 10 & 6.84 & 966 & 98 & 5.36 & 34 & 5.59 & 1162 & 160 & 13.9 & 83 & 7.33 & 165 & 116 & 4.36 & 72 & 3.23 & 171 & 3.82 & 122 & 3.06 & 1246 & 2770 & 1570 & 418 & 42 & 340 & 1600 & 34495 \\
Huawei Nova 5 & HiSilicon Kirin 810 & 39 & 25 & 6.99 & 10 & 6.84 & 923 & 98 & 5.36 & 34 & 5.59 & 1163 & 160 & 13.9 & 82 & 7.33 & 226 & 115 & 4.36 & 72 & 3.23 & 170 & 3.82 & 122 & 3.06 & 1278 & 2818 & 1586 & 423 & 42 & 339 & 1600 & 34432 \\
Asus ROG Phone II & Snapdragon 855 Plus & 42 & 4.9 & 11.65 & 16 & 7.34 & 354 & 16 & 31.88 & 57 & 26.43 & 393 & 40 & 16.66 & 142 & 21.66 & 82 & 24 & 10.02 & 66 & 39.65 & 45 & 5.34 & 183 & 4.11 & 585 & 1583 & 1215 & 142 & 67 & 115 & 1000 & 32727 \\
Asus Zenfone 6 & Snapdragon 855 & 64 & 8.3 & 11.44 & 25 & 6.88 & 414 & 18 & 31.59 & 70 & 23.65 & 379 & 47 & 14.62 & 169 & 12.87 & 88 & 30 & 5.8 & 87 & 37.92 & 48 & 3.61 & 210 & 3.07 & 653 & 1673 & 1361 & 214 & 83 & 135 & 1000 & 27410 \\
Samsung Galaxy Note10 & Snapdragon 855 & 50 & 8.2 & 11.65 & 25 & 7.34 & 402 & 19 & 31.87 & 70 & 26.43 & 443 & 45 & 16.66 & 164 & 21.66 & 87 & 29 & 10.02 & 84 & 39.65 & 47 & 5.34 & 215 & 4.11 & 587 & 1636 & 1332 & 165 & 88 & 133 & 1000 & 27151 \\
Oppo Reno Z & MediaTek Helio P90 & 71 & 4 & 6.77 & 8.3 & 6.72 & 774 & 23 & 5.33 & 101 & 4.87 & 962 & 38 & 6 & 263 & 5.91 & 169 & 22 & 3.78 & 75 & 4.31 & 147 & 3.45 & 309 & 3.06 & 1321 & 3489 & 2080 & 2187 & 66 & 1236 & 1000 & 26738 \\
Sony Xperia 1 & Snapdragon 855 & 90 & 9.3 & 11.44 & 24 & 6.88 & 428 & 20 & 31.57 & 74 & 23.65 & 407 & 46 & 14.62 & 177 & 12.87 & 87 & 29 & 5.8 & 86 & 37.92 & 48 & 3.61 & 212 & 3.07 & 594 & 1669 & 1374 & 182 & 88 & 143 & 1000 & 26672 \\
LG G8S ThinQ & Snapdragon 855 & 91 & 8.7 & 11.44 & 24 & 6.88 & 445 & 19 & 31.57 & 73 & 23.65 & 441 & 45 & 14.62 & 175 & 12.87 & 88 & 28 & 5.8 & 86 & 37.92 & 47 & 3.61 & 211 & 3.07 & 671 & 1789 & 1488 & 221 & 86 & 143 & 1000 & 26493 \\
Xiaomi Mi 9T Pro & Snapdragon 855 & 85 & 7.5 & 11.65 & 21 & 7.34 & 474 & 18 & 31.8 & 68 & 26.43 & 438 & 44 & 16.66 & 165 & 21.66 & 87 & 28 & 10.02 & 83 & 39.65 & 47 & 5.34 & 208 & 4.11 & 583 & 1618 & 1399 & 272 & 89 & 158 & 1000 & 26257 \\
Oppo Reno 10x zoom & Snapdragon 855 & 88 & 9 & 11.44 & 23 & 6.88 & 563 & 19 & 31.61 & 70 & 23.65 & 576 & 44 & 14.62 & 166 & 12.87 & 99 & 29 & 5.8 & 83 & 37.92 & 48 & 3.61 & 232 & 3.07 & 691 & 1717 & 1550 & 189 & 81 & 146 & 1000 & 26144 \\
Xiaomi Redmi K20 Pro & Snapdragon 855 & 94 & 7.5 & 11.65 & 22 & 7.34 & 526 & 18 & 31.84 & 67 & 26.43 & 487 & 44 & 16.66 & 164 & 21.66 & 98 & 28 & 10.02 & 82 & 39.65 & 47 & 5.34 & 204 & 4.11 & 759 & 1681 & 1713 & 187 & 87 & 138 & 1000 & 25867 \\
OnePlus 7 & Snapdragon 855 & 91 & 8.4 & 11.65 & 22 & 7.34 & 429 & 18 & 31.86 & 70 & 26.43 & 443 & 44 & 16.66 & 166 & 21.66 & 87 & 29 & 10.02 & 84 & 39.65 & 47 & 5.34 & 210 & 4.11 & 920 & 1920 & 1362 & 215 & 84 & 138 & 1000 & 25804 \\
OnePlus 7 Pro & Snapdragon 855 & 91 & 8.8 & 11.65 & 23 & 7.34 & 426 & 19 & 31.88 & 72 & 26.43 & 415 & 44 & 16.66 & 172 & 21.66 & 88 & 28 & 10.02 & 85 & 39.65 & 47 & 5.34 & 212 & 4.11 & 771 & 1889 & 1374 & 202 & 85 & 145 & 1000 & 25720 \\
Samsung Galaxy Note10 & Exynos 9825 Octa & 32 & 19 & 6.99 & 17 & 7.02 & 458 & 63 & 11.11 & 65 & 27.14 & 784 & 128 & 9.42 & 151 & 15.9 & 278 & 75 & 7.97 & 124 & 9.79 & 199 & 5.47 & 158 & 4.1 & 820 & 1775 & 1240 & 203 & 67 & 200 & 2000 & 25470 \\
Lenovo Z6 Pro & Snapdragon 855 & 94 & 9.4 & 11.44 & 25 & 6.88 & 451 & 19 & 31.46 & 73 & 23.65 & 447 & 45 & 14.62 & 182 & 12.87 & 87 & 30 & 5.8 & 88 & 37.92 & 47 & 3.61 & 214 & 3.07 & 878 & 1887 & 1384 & 243 & 89 & 175 & 1000 & 25268 \\
Samsung Galaxy S10+ & Snapdragon 855 & 51 & 8.5 & 11.44 & 25 & 6.88 & 450 & 19 & 31.62 & 69 & 23.65 & 445 & 44 & 14.62 & 164 & 12.87 & 87 & 33 & 5.8 & 84 & 37.92 & 451 & 3.58 & 213 & 3.07 & 618 & 1652 & 1392 & 171 & 84 & 134 & 1000 & 25087 \\
Samsung Galaxy S10 & Snapdragon 855 & 52 & 8.9 & 11.44 & 26 & 6.88 & 458 & 20 & 31.61 & 69 & 23.65 & 446 & 45 & 14.62 & 167 & 12.87 & 87 & 33 & 5.8 & 85 & 37.92 & 452 & 3.58 & 216 & 3.07 & 641 & 1687 & 1396 & 177 & 88 & 134 & 1000 & 24646 \\
Samsung Galaxy S10e & Snapdragon 855 & 51 & 8.8 & 11.44 & 27 & 6.88 & 451 & 20 & 31.47 & 70 & 23.65 & 446 & 45 & 14.62 & 166 & 12.87 & 87 & 33 & 5.8 & 84 & 37.92 & 451 & 3.58 & 214 & 3.07 & 647 & 1685 & 1396 & 199 & 87 & 135 & 1000 & 24518 \\
Xiaomi Mi 9 Explorer & Snapdragon 855 & 80 & 8.1 & 11.65 & 21 & 7.34 & 473 & 19 & 31.89 & 60 & 26.43 & 513 & 45 & 16.66 & 156 & 21.66 & 99 & 32 & 10.02 & 75 & 39.65 & 450 & 5.33 & 187 & 4.11 & 781 & 1967 & 1673 & 208 & 79 & 138 & 1000 & 24241 \\
Huawei P30 Pro & HiSilicon Kirin 980 & 51 & 79 & 6.73 & 26 & 6.61 & 523 & 297 & 5.19 & 59 & 6.24 & 494 & 707 & 628 & 118 & 17.52 & 82 & 395 & 3.57 & 102 & 3.14 & 1114 & 3.45 & 164 & 2.97 & 814 & 1771 & 1330 & 2430 & 77 & 1109 & 2000 & 23874 \\
LG G8 ThinQ & Snapdragon 855 & 96 & 9.5 & 11.44 & 24 & 6.88 & 414 & 19 & 31.5 & 73 & 23.65 & 384 & 47 & 14.62 & 186 & 12.87 & 89 & 33 & 5.8 & 87 & 37.92 & 454 & 3.58 & 215 & 3.07 & 591 & 1642 & 1389 & 210 & 89 & 156 & 1000 & 23499 \\
Xiaomi Mi 9 & Snapdragon 855 & 92 & 8.8 & 11.65 & 23 & 7.34 & 439 & 19 & 31.85 & 70 & 26.43 & 431 & 45 & 16.66 & 166 & 21.66 & 90 & 34 & 10.02 & 86 & 39.65 & 453 & 5.33 & 211 & 4.11 & 587 & 1643 & 1402 & 231 & 89 & 142 & 1000 & 23199 \\
Huawei Mate 20 Pro & HiSilicon Kirin 980 & 52 & 90 & 6.73 & 21 & 6.6 & 553 & 299 & 5.19 & 55 & 8.77 & 519 & 743 & 628 & 114 & 29.45 & 85 & 380 & 3.57 & 83 & 3.14 & 1084 & 3.45 & 12 & 484 & 802 & 1795 & 1327 & 2380 & 56 & 1150 & 2000 & 21125 \\
Huawei Mate 20 & HiSilicon Kirin 980 & 52 & 88 & 6.73 & 21 & 6.6 & 540 & 307 & 5.19 & 53 & 8.77 & 491 & 744 & 628 & 114 & 29.45 & 86 & 378 & 3.57 & 90 & 3.14 & 1085 & 3.45 & 12 & 484 & 800 & 1798 & 1331 & 2311 & 58 & 1178 & 2000 & 20973 \\
Huawei Mate 20 X & HiSilicon Kirin 980 & 52 & 90 & 6.73 & 21 & 6.6 & 554 & 295 & 5.19 & 53 & 8.77 & 505 & 734 & 628 & 114 & 29.45 & 83 & 381 & 3.57 & 88 & 3.14 & 1086 & 3.45 & 13 & 484 & 798 & 1799 & 1330 & 1999 & 56 & 1173 & 2000 & 20959 \\
Honor View 20 & HiSilicon Kirin 980 & 52 & 85 & 6.73 & 22 & 6.6 & 518 & 308 & 5.19 & 53 & 8.77 & 505 & 720 & 628 & 113 & 29.45 & 90 & 397 & 3.57 & 86 & 3.14 & 1102 & 3.45 & 14 & 484 & 800 & 1798 & 1343 & 2297 & 57 & 1177 & 2000 & 20674 \\
Samsung Galaxy S9+ & Snapdragon 845 & 108 & 11 & 11.44 & 25 & 6.8 & 524 & 44 & 19.37 & 81 & 24.95 & 452 & 91 & 14.62 & 202 & 12.07 & 165 & 70 & 5.8 & 98 & 37.92 & 610 & 3.58 & 262 & 3.07 & 918 & 2495 & 1759 & 213 & 93 & 169 & 1000 & 18885 \\
vivo NEX Dual Display & Snapdragon 845 & 122 & 11 & 11.44 & 28 & 6.8 & 534 & 45 & 19.37 & 80 & 24.95 & 453 & 92 & 14.62 & 203 & 12.07 & 161 & 70 & 5.8 & 96 & 37.92 & 621 & 3.58 & 261 & 3.07 & 772 & 2297 & 1751 & 232 & 93 & 167 & 1000 & 18710 \\
Samsung Galaxy S9 & Snapdragon 845 & 101 & 11 & 11.44 & 24 & 6.8 & 573 & 46 & 19.37 & 80 & 24.95 & 535 & 92 & 14.62 & 207 & 12.07 & 165 & 71 & 5.8 & 100 & 37.92 & 611 & 3.58 & 264 & 3.07 & 927 & 2452 & 1782 & 217 & 95 & 169 & 1000 & 18591 \\
Samsung Galaxy Note9 & Snapdragon 845 & 109 & 11 & 11.44 & 27 & 6.8 & 538 & 45 & 19.37 & 81 & 24.95 & 493 & 91 & 14.62 & 204 & 12.07 & 165 & 71 & 5.8 & 100 & 37.92 & 610 & 3.58 & 265 & 3.07 & 973 & 2566 & 1759 & 209 & 93 & 168 & 1000 & 18509 \\
LG G7 ThinQ & Snapdragon 845 & 124 & 11 & 11.44 & 28 & 6.8 & 533 & 44 & 19.37 & 81 & 24.95 & 474 & 90 & 14.62 & 203 & 12.07 & 168 & 70 & 5.8 & 96 & 37.92 & 609 & 3.58 & 262 & 3.07 & 988 & 2812 & 1865 & 232 & 94 & 168 & 1000 & 18306 \\
Asus Zenfone 5z & Snapdragon 845 & 65 & 8.5 & 11.65 & 16 & 6.62 & 523 & 45 & 17.97 & 147 & 11.13 & 465 & 88 & 16.66 & 330 & 15.32 & 159 & 77 & 10.02 & 236 & 9.76 & 742 & 5.33 & 657 & 4.16 & 822 & 2393 & 1715 & 208 & 186 & 186 & 1000 & 16450 \\
OnePlus 6 & Snapdragon 845 & 118 & 11 & 11.65 & 28 & 6.62 & 537 & 70 & 17.85 & 159 & 11.13 & 457 & 90 & 16.66 & 348 & 15.32 & 166 & 96 & 10.02 & 312 & 9.76 & 692 & 5.33 & 692 & 4.16 & 819 & 2341 & 1735 & 252 & 220 & 213 & 1000 & 14113 \\
OnePlus 6T & Snapdragon 845 & 119 & 12 & 11.65 & 28 & 6.62 & 538 & 71 & 17.85 & 160 & 11.13 & 457 & 90 & 16.66 & 348 & 15.32 & 167 & 97 & 10.02 & 314 & 9.76 & 693 & 5.33 & 693 & 4.16 & 817 & 2322 & 1717 & 258 & 220 & 214 & 1000 & 14054 \\
Xiaomi Mi 9T & Snapdragon 730 & 81 & 11 & 11.44 & 29 & 6.88 & 814 & 44 & 19.37 & 160 & 23.65 & 1069 & 89 & 14.62 & 428 & 12.87 & 123 & 66 & 5.8 & 198 & 37.92 & 110 & 3.61 & 619 & 3.07 & 1396 & 3074 & 1658 & 445 & 187 & 378 & 1000 & 13977 \\
Xiaomi Redmi K20 & Snapdragon 730 & 74 & 11 & 11.44 & 28 & 6.88 & 876 & 44 & 19.37 & 160 & 23.65 & 1099 & 88 & 14.62 & 424 & 12.87 & 127 & 66 & 5.8 & 195 & 37.92 & 110 & 3.61 & 618 & 3.07 & 1439 & 3188 & 1722 & 435 & 184 & 391 & 1000 & 13947 \\
Samsung Galaxy A80 & Snapdragon 730 & 90 & 13 & 11.44 & 31 & 6.88 & 846 & 47 & 19.37 & 150 & 23.65 & 1108 & 90 & 14.62 & 391 & 12.87 & 126 & 69 & 5.8 & 185 & 37.92 & 111 & 3.61 & 553 & 3.07 & 1453 & 3093 & 1645 & 406 & 175 & 347 & 1000 & 13940 \\
Lenovo Z6 & Snapdragon 730 & 90 & 11 & 11.44 & 29 & 6.88 & 808 & 45 & 19.37 & 163 & 23.65 & 1057 & 91 & 14.62 & 426 & 12.87 & 127 & 66 & 5.8 & 199 & 37.92 & 111 & 3.61 & 617 & 3.07 & 1809 & 3651 & 1643 & 427 & 181 & 378 & 1000 & 13571 \\
Xiaomi Red. Note 8 Pro & MediaTek G90T & 41 & 15 & 6.77 & 37 & 6.92 & 607 & 64 & 5.31 & 223 & 20.13 & 947 & 139 & 5.91 & 584 & 10.55 & 92 & 107 & 3.78 & 459 & 4.35 & 308 & 3.45 & 977 & 3.08 & 1276 & 2796 & 1672 & 482 & 665 & 1057 & 1400 & 12574 \\
Meizu 16Xs & Snapdragon 675 & 83 & 11 & 11.44 & 40 & 6.88 & 852 & 34 & 31.54 & 314 & 25.83 & 1130 & 74 & 14.62 & 898 & 12.87 & 125 & 54 & 5.8 & 526 & 38.02 & 102 & 3.61 & 1253 & 3.07 & 1514 & 3219 & 1788 & 844 & 351 & 729 & 1000 & 11394 \\
Samsung Galaxy S10+ & Exynos 9820 Octa & 36 & 28 & 6.73 & 41 & 6.6 & 490 & 96 & 5.19 & 186 & 5.31 & 501 & 197 & 6.4 & 431 & 6.27 & 284 & 410 & 3.57 & 897 & 3.22 & 445 & 3.45 & 831 & 3.11 & 799 & 1856 & 1283 & 1063 & 838 & 841 & 500 & 10315 \\
Samsung Galaxy S10e & Exynos 9820 Octa & 35 & 26 & 6.73 & 38 & 6.6 & 486 & 97 & 5.19 & 185 & 5.31 & 479 & 211 & 6.4 & 471 & 6.27 & 288 & 403 & 3.57 & 876 & 3.22 & 446 & 3.45 & 861 & 3.11 & 818 & 1891 & 1305 & 1059 & 831 & 834 & 500 & 10296 \\
Samsung Galaxy A70 & Snapdragon 675 & 88 & 13 & 11.44 & 43 & 6.88 & 836 & 41 & 31.52 & 342 & 25.83 & 1137 & 90 & 14.62 & 967 & 12.87 & 130 & 67 & 5.8 & 649 & 38.02 & 134 & 3.61 & 1368 & 3.07 & 1506 & 3180 & 1783 & 878 & 376 & 779 & 1000 & 10246 \\
Samsung Galaxy S10 & Exynos 9820 Octa & 35 & 27 & 6.73 & 38 & 6.6 & 484 & 96 & 5.19 & 187 & 5.31 & 494 & 209 & 6.4 & 462 & 6.27 & 294 & 401 & 3.57 & 875 & 3.22 & 447 & 3.45 & 849 & 3.11 & 822 & 1902 & 1387 & 1081 & 833 & 843 & 500 & 10221 \\
Huawei Mate 10 Pro & HiSilicon Kirin 970 & 93 & 157 & 6.73 & 43 & 172 & 732 & 371 & 5.19 & 69 & 58.21 & 587 & 762 & 628 & 1457 & 6.27 & 174 & 506 & 3.57 & 138 & 3.2 & 1509 & 3.45 & 231 & 2.98 & 992 & 3037 & 2494 & 2766 & 79 & 1181 & 600 & 9064 \\
Huawei P20 Pro & HiSilicon Kirin 970 & 93 & 133 & 6.73 & 44 & 172 & 730 & 382 & 5.19 & 69 & 58.21 & 585 & 757 & 628 & 1403 & 6.27 & 174 & 512 & 3.57 & 147 & 3.2 & 1505 & 3.45 & 238 & 2.98 & 965 & 2987 & 2496 & 3069 & 83 & 1130 & 600 & 9005 \\
Honor Play & HiSilicon Kirin 970 & 93 & 148 & 6.73 & 43 & 172 & 731 & 383 & 5.19 & 68 & 58.21 & 595 & 802 & 628 & 1636 & 6.27 & 175 & 536 & 3.57 & 138 & 3.2 & 1534 & 3.45 & 230 & 2.98 & 1068 & 3128 & 2495 & 3083 & 78 & 1239 & 600 & 8919 \\
Huawei Honor 10 & HiSilicon Kirin 970 & 94 & 142 & 6.73 & 43 & 172 & 736 & 417 & 5.19 & 67 & 58.21 & 601 & 775 & 628 & 1603 & 6.27 & 175 & 536 & 3.57 & 130 & 3.2 & 1529 & 3.45 & 227 & 2.98 & 1120 & 3218 & 2494 & 2904 & 80 & 1258 & 600 & 8906 \\
Huawei P20 & HiSilicon Kirin 970 & 94 & 135 & 6.73 & 43 & 172 & 728 & 360 & 5.19 & 68 & 58.21 & 593 & 779 & 628 & 1409 & 6.27 & 173 & 550 & 3.57 & 151 & 3.2 & 1523 & 3.45 & 246 & 2.98 & 983 & 2978 & 2496 & 3276 & 92 & 1160 & 600 & 8892 \\
Huawei Honor View 10 & HiSilicon Kirin 970 & 94 & 127 & 6.73 & 43 & 172 & 730 & 402 & 5.19 & 72 & 58.21 & 587 & 825 & 628 & 1799 & 6.27 & 175 & 499 & 3.57 & 132 & 3.2 & 1498 & 3.45 & 224 & 2.98 & 1081 & 3186 & 2493 & 2362 & 93 & 1246 & 600 & 8732 \\
Xiaomi Mi A3 & Snapdragon 665 & 138 & 13 & 11.44 & 50 & 6.88 & 894 & 52 & 31.53 & 483 & 25.83 & 708 & 118 & 14.62 & 1292 & 12.87 & 212 & 94 & 5.8 & 678 & 38.02 & 192 & 3.61 & 2174 & 3.07 & 1165 & 3630 & 3095 & 1292 & 553 & 1149 & 1000 & 8187 \\
Google Pixel 3 XL & Snapdragon 845 & 84 & 10 & 11.44 & 159 & 6.6 & 542 & 70 & 19.37 & 731 & 5.31 & 422 & 92 & 14.62 & 1384 & 6.27 & 185 & 94 & 5.8 & 1514 & 3.22 & 692 & 3.58 & 3479 & 3.11 & 828 & 2897 & 2084 & 3173 & 1223 & 1203 & 400 & 7999 \\
Google Pixel 3 & Snapdragon 845 & 87 & 11 & 11.44 & 139 & 6.6 & 535 & 69 & 19.37 & 695 & 5.31 & 421 & 93 & 14.62 & 1373 & 6.27 & 186 & 94 & 5.8 & 1541 & 3.22 & 692 & 3.58 & 3524 & 3.11 & 793 & 2753 & 2180 & 3322 & 1246 & 1220 & 400 & 7977 \\
Samsung Galaxy Note9 & Exynos 9810 Octa & 99 & 45 & 6.73 & 72 & 6.6 & 604 & 166 & 5.19 & 209 & 5.31 & 688 & 360 & 6.4 & 488 & 6.27 & 220 & 539 & 3.57 & 1574 & 3.22 & 852 & 3.45 & 843 & 3.11 & 1083 & 1753 & 1476 & 1490 & 787 & 779 & 500 & 7937 \\
Xiaomi Mi CC9e & Snapdragon 665 & 137 & 14 & 11.65 & 49 & 7.34 & 902 & 53 & 31.81 & 482 & 27.45 & 709 & 119 & 16.66 & 1272 & 21.66 & 213 & 94 & 10.02 & 674 & 39.74 & 192 & 5.34 & 2169 & 4.11 & 1183 & 3643 & 3101 & 1289 & 550 & 1135 & 1000 & 7935 \\
Xiaomi Mi 8 & Snapdragon 845 & 113 & 11 & 11.65 & 125 & 6.62 & 566 & 68 & 17.85 & 725 & 11.13 & 493 & 90 & 16.66 & 1428 & 15.32 & 172 & 94 & 10.02 & 1532 & 9.76 & 690 & 5.33 & 3269 & 4.16 & 976 & 2814 & 2024 & 2944 & 1323 & 1220 & 500 & 7695 \\
Xiaomi Pocophone F1 & Snapdragon 845 & 122 & 11 & 11.65 & 119 & 6.62 & 566 & 67 & 17.85 & 727 & 11.13 & 502 & 90 & 16.66 & 1416 & 15.32 & 175 & 92 & 10.02 & 1523 & 9.76 & 687 & 5.33 & 3220 & 4.16 & 1077 & 2934 & 2021 & 3037 & 1282 & 1216 & 500 & 7557 \\
vivo V15 & MediaTek Helio P70 & 106 & 26 & 6.77 & 66 & 6.99 & 805 & 89 & 5.31 & 374 & 19.98 & 667 & 181 & 5.91 & 932 & 9.81 & 191 & 163 & 3.78 & 1096 & 4.42 & 474 & 3.45 & 865 & 3.09 & 1158 & 3547 & 2782 & 759 & 764 & 1202 & 500 & 7512 \\
Xiaomi Mi Mix 3 & Snapdragon 845 & 113 & 12 & 11.65 & 118 & 6.62 & 577 & 72 & 17.85 & 685 & 11.13 & 526 & 90 & 16.66 & 1354 & 15.32 & 185 & 96 & 10.02 & 1613 & 9.76 & 691 & 5.33 & 3190 & 4.16 & 982 & 2865 & 2326 & 3104 & 1346 & 1174 & 500 & 7402 \\
Xiaomi Mi Mix 2S & Snapdragon 845 & 118 & 11 & 11.65 & 137 & 6.62 & 590 & 67 & 17.85 & 810 & 11.13 & 515 & 89 & 16.66 & 1587 & 15.32 & 181 & 92 & 10.02 & 1570 & 9.76 & 686 & 5.33 & 3335 & 4.16 & 1060 & 2913 & 2238 & 2964 & 1399 & 1319 & 500 & 7365 \\
Lenovo Z6 Youth & Snapdragon 710 & 132 & 11 & 11.44 & 155 & 6.6 & 1083 & 48 & 19.37 & 924 & 5.31 & 1300 & 89 & 14.62 & 1849 & 6.27 & 218 & 66 & 5.8 & 1994 & 3.22 & 110 & 3.61 & 4632 & 3.11 & 1895 & 4666 & 2638 & 3499 & 1696 & 1604 & 500 & 7331 \\
Meizu Note 9 & Snapdragon 675 & 93 & 11 & 6.87 & 134 & 6.6 & 857 & 62 & 5.42 & 769 & 5.31 & 1123 & 147 & 7.09 & 1741 & 6.27 & 133 & 90 & 4.92 & 1834 & 3.22 & 830 & 3.58 & 4742 & 3.11 & 1521 & 3248 & 1798 & 3449 & 1486 & 1441 & 500 & 7075 \\
Samsung Galaxy S9+ & Exynos 9810 Octa & 119 & 46 & 6.73 & 75 & 6.6 & 1080 & 198 & 5.19 & 241 & 5.31 & 722 & 395 & 6.4 & 531 & 6.27 & 253 & 593 & 3.57 & 1636 & 3.22 & 885 & 3.45 & 871 & 3.11 & 1138 & 2233 & 1860 & 1515 & 792 & 791 & 500 & 6914 \\
Samsung Galaxy S9 & Exynos 9810 Octa & 121 & 47 & 6.73 & 74 & 6.6 & 926 & 179 & 5.19 & 217 & 5.31 & 741 & 376 & 6.4 & 504 & 6.27 & 262 & 600 & 3.57 & 1646 & 3.22 & 898 & 3.45 & 885 & 3.11 & 1160 & 2202 & 1871 & 1530 & 794 & 802 & 400 & 6825 \\
Xiaomi Red. Note 7 Pro & Snapdragon 675 & 79 & 11 & 7.12 & 405 & 6.62 & 893 & 62 & 9.75 & 908 & 11.13 & 1167 & 146 & 9.66 & 1693 & 15.32 & 130 & 89 & 7.97 & 1871 & 9.76 & 833 & 5.34 & 4731 & 4.16 & 1588 & 3448 & 1858 & 3159 & 1487 & 1450 & 500 & 6702 \\
vivo V15 Pro & Snapdragon 675 & 72 & 12 & 6.87 & 151 & 6.6 & 934 & 62 & 5.42 & 1239 & 5.31 & 1644 & 154 & 7.09 & 2928 & 6.27 & 128 & 89 & 4.92 & 2227 & 3.22 & 834 & 3.58 & 6485 & 3.11 & 1613 & 3599 & 1762 & 3107 & 1925 & 1904 & 500 & 6687 \\
vivo S1 & MediaTek Helio P65 & 79 & 43 & 6.77 & 54 & 6.92 & 934 & 254 & 6.2 & 347 & 20.13 & 1169 & 529 & 5.91 & 921 & 10.55 & 190 & 654 & 3.78 & 748 & 4.35 & 1167 & 3.45 & 1672 & 3.08 & 1466 & 3969 & 2309 & 960 & 954 & 1439 & 1000 & 6643 \\
Xiaomi Mi 9 SE & Snapdragon 712 & 132 & 12 & 11.44 & 193 & 6.6 & 990 & 47 & 19.37 & 838 & 5.31 & 1266 & 95 & 14.62 & 1604 & 6.27 & 205 & 69 & 5.8 & 1838 & 3.22 & 608 & 3.58 & 3922 & 3.11 & 1609 & 3905 & 2298 & 3383 & 1446 & 1451 & 500 & 6556 \\
vivo X27 & Snapdragon 710 & 131 & 12 & 11.44 & 154 & 6.6 & 1011 & 46 & 19.37 & 838 & 5.31 & 1269 & 97 & 14.62 & 1828 & 6.27 & 205 & 68 & 5.8 & 2018 & 3.22 & 607 & 3.58 & 4416 & 3.11 & 1633 & 4011 & 2344 & 3247 & 1594 & 1412 & 500 & 6505 \\
vivo X27 Pro & Snapdragon 710 & 133 & 12 & 11.44 & 143 & 6.6 & 1010 & 47 & 19.37 & 876 & 5.31 & 1289 & 96 & 14.62 & 1880 & 6.27 & 205 & 71 & 5.8 & 1960 & 3.22 & 607 & 3.58 & 4471 & 3.11 & 1701 & 4021 & 2354 & 3869 & 1503 & 1491 & 500 & 6474 \\
Google Pixel 3a XL & Snapdragon 670 & 87 & 13 & 11.44 & 31 & 6.88 & 854 & 47 & 19.37 & 149 & 23.65 & 1105 & 92 & 14.62 & 390 & 12.87 & 126 & 69 & 5.8 & 184 & 37.92 & 111 & 3.61 & 554 & 3.07 & 1475 & 4125 & 1665 & 407 & 173 & 341 & 400 & 6444 \\
Xiaomi Mi 8 SE & Snapdragon 710 & 132 & 12 & 11.44 & 179 & 6.6 & 1037 & 46 & 19.37 & 866 & 5.31 & 1283 & 96 & 14.62 & 2088 & 6.27 & 210 & 70 & 5.8 & 1914 & 3.22 & 608 & 3.58 & 4180 & 3.11 & 1706 & 4120 & 2504 & 4481 & 1683 & 1590 & 500 & 6355 \\
Oppo Reno & Snapdragon 710 & 133 & 12 & 11.44 & 211 & 6.6 & 1103 & 48 & 19.37 & 838 & 5.31 & 1302 & 95 & 14.62 & 1598 & 6.27 & 239 & 70 & 5.8 & 1844 & 3.22 & 603 & 3.58 & 3812 & 3.11 & 1589 & 4052 & 2926 & 3486 & 1493 & 1380 & 500 & 6354 \\
Realme 3 & MediaTek Helio P70 & 111 & 28 & 6.77 & 69 & 6.99 & 864 & 92 & 5.31 & 504 & 19.98 & 731 & 185 & 5.91 & 1172 & 9.81 & 211 & 165 & 3.78 & 1466 & 4.42 & 484 & 3.45 & 1249 & 3.09 & 1246 & 3649 & 2809 & 1261 & 1405 & 1832 & 400 & 6330 \\
Oppo F11 Pro & MediaTek Helio P70 & 109 & 32 & 6.77 & 66 & 6.99 & 840 & 143 & 5.31 & 479 & 19.98 & 715 & 320 & 5.91 & 966 & 9.81 & 191 & 314 & 3.78 & 1778 & 4.31 & 778 & 3.45 & 1080 & 3.09 & 1118 & 3368 & 2791 & 1945 & 1303 & 1257 & 500 & 6301 \\
Realme 3 Pro & Snapdragon 710 & 134 & 12 & 11.44 & 215 & 6.6 & 1099 & 47 & 19.37 & 897 & 5.31 & 1294 & 94 & 14.62 & 1706 & 6.27 & 242 & 68 & 5.8 & 1926 & 3.22 & 608 & 3.58 & 4015 & 3.11 & 1660 & 4036 & 2908 & 3567 & 1623 & 1401 & 500 & 6269 \\
Oppo K3 & Snapdragon 710 & 131 & 13 & 11.44 & 215 & 6.6 & 1099 & 47 & 19.37 & 891 & 5.31 & 1300 & 94 & 14.62 & 1649 & 6.27 & 244 & 69 & 5.8 & 1971 & 3.22 & 602 & 3.58 & 3966 & 3.11 & 1642 & 4044 & 2920 & 3392 & 1567 & 1405 & 500 & 6241 \\
Nokia X7 & Snapdragon 710 & 132 & 11 & 11.65 & 148 & 6.62 & 1020 & 49 & 17.97 & 904 & 11.13 & 1233 & 95 & 16.66 & 1914 & 15.32 & 205 & 80 & 10.02 & 2030 & 9.76 & 737 & 5.33 & 4679 & 4.16 & 1484 & 3942 & 2358 & 3135 & 1644 & 1570 & 500 & 6119 \\
Lenovo Z5s & Snapdragon 710 & 132 & 10 & 11.44 & 143 & 6.6 & 1006 & 47 & 19.33 & 1005 & 5.31 & 1351 & 95 & 14.62 & 1943 & 6.27 & 211 & 80 & 5.8 & 2148 & 3.22 & 737 & 3.58 & 5570 & 3.11 & 2225 & 4680 & 2314 & 3422 & 2127 & 1749 & 500 & 6060 \\
Oppo F7 Youth & MediaTek Helio P60 & 113 & 31 & 6.77 & 66 & 6.99 & 855 & 143 & 5.31 & 461 & 19.98 & 738 & 319 & 5.91 & 1036 & 9.81 & 201 & 314 & 3.78 & 1806 & 4.31 & 785 & 3.45 & 2927 & 3.1 & 1153 & 3543 & 2957 & 2472 & 1290 & 1322 & 500 & 5921 \\
Oppo F11 & MediaTek Helio P70 & 108 & 32 & 6.77 & 69 & 6.99 & 836 & 144 & 5.31 & 489 & 19.98 & 728 & 321 & 5.91 & 1051 & 9.81 & 190 & 309 & 3.78 & 1784 & 4.31 & 786 & 3.45 & 1172 & 3.09 & 1101 & 3372 & 2812 & 7102 & 1293 & 1340 & 300 & 5763 \\
Motorola One Action & Exynos 9609 & 131 & 49 & 6.73 & 64 & 6.6 & 862 & 388 & 5.19 & 445 & 5.31 & 674 & 942 & 6.4 & 1233 & 6.27 & 197 & 1024 & 3.57 & 1660 & 3.22 & 2330 & 3.45 & 1490 & 3.11 & 1136 & 3930 & 2841 & 1738 & 732 & 762 & 500 & 5730 \\
Motorola One Vision & Exynos 9609 & 129 & 50 & 6.73 & 61 & 6.6 & 870 & 383 & 5.19 & 444 & 5.31 & 672 & 937 & 6.4 & 1230 & 6.27 & 201 & 1027 & 3.57 & 1661 & 3.22 & 2325 & 3.45 & 1448 & 3.11 & 1136 & 4269 & 2912 & 1828 & 731 & 735 & 400 & 5669 \\
Sony Xperia XZ3 & Snapdragon 845 & 121 & 94 & 6.99 & 159 & 6.62 & 538 & 398 & 10.57 & 712 & 11.13 & 474 & 920 & 629 & 1462 & 15.32 & 167 & 416 & 7.97 & 1557 & 9.76 & 1605 & 5.47 & 3736 & 4.16 & 1339 & 3097 & 1850 & 3493 & 1274 & 1095 & 400 & 5503 \\
Samsung Galaxy A50 & Exynos 9610 & 157 & 53 & 6.73 & 78 & 6.6 & 1116 & 382 & 5.19 & 460 & 5.31 & 665 & 931 & 6.4 & 1238 & 6.27 & 187 & 1023 & 3.57 & 1661 & 3.22 & 2323 & 3.45 & 1456 & 3.11 & 1216 & 4091 & 2837 & 1869 & 769 & 713 & 500 & 5399 \\
Nokia 9 PureView & Snapdragon 845 & 123 & 381 & 6.99 & 452 & 6.62 & 543 & 1233 & 10.57 & 5987 & 11.13 & 501 & 2125 & 629 & 7087 & 15.32 & 166 & 538 & 7.97 & 1888 & 9.76 & 1771 & 5.47 & 4975 & 4.16 & 891 & 2462 & 1839 & 4838 & 1735 & 1703 & 500 & 5223 \\
Samsung Galaxy Note8 & Snapdragon 835 & 136 & 101 & 6.73 & 154 & 6.6 & 804 & 378 & 5.19 & 792 & 5.31 & 636 & 770 & 628 & 1534 & 6.27 & 187 & 515 & 3.57 & 1665 & 3.22 & 1363 & 3.45 & 3732 & 3.11 & 1037 & 3409 & 2922 & 2851 & 1387 & 1377 & 500 & 5059 \\
HTC U11 & Snapdragon 835 & 138 & 102 & 6.73 & 154 & 6.6 & 767 & 373 & 5.19 & 768 & 5.31 & 628 & 790 & 628 & 1500 & 6.27 & 186 & 583 & 3.57 & 1673 & 3.22 & 1656 & 3.45 & 3968 & 3.11 & 1130 & 3479 & 2904 & 3194 & 1329 & 1289 & 500 & 5039 \\
Essential Phone & Snapdragon 835 & 140 & 102 & 6.73 & 149 & 6.6 & 820 & 358 & 5.19 & 749 & 5.31 & 638 & 738 & 628 & 1495 & 6.27 & 184 & 551 & 3.57 & 1782 & 3.22 & 1413 & 3.45 & 3727 & 3.11 & 1032 & 3326 & 2827 & 2871 & 1362 & 1295 & 500 & 5009 \\
Google Pixel 2 & Snapdragon 835 & 130 & 193 & 6.73 & 204 & 6.6 & 746 & 467 & 5.19 & 891 & 5.31 & 651 & 835 & 628 & 1473 & 6.27 & 215 & 632 & 3.57 & 1700 & 3.22 & 1659 & 3.45 & 3665 & 3.11 & 1117 & 3470 & 2570 & 3157 & 1290 & 1156 & 400 & 4859 \\
Google Pixel 2 XL & Snapdragon 835 & 130 & 202 & 6.73 & 208 & 6.6 & 742 & 426 & 5.19 & 875 & 5.31 & 653 & 848 & 628 & 1499 & 6.27 & 216 & 639 & 3.57 & 1704 & 3.22 & 1696 & 3.45 & 3743 & 3.11 & 1194 & 3585 & 2567 & 3147 & 1289 & 1173 & 500 & 4851 \\
Google Pixel XL & Snapdragon 821 & 109 & 105 & 6.73 & 125 & 6.6 & 761 & 506 & 5.19 & 956 & 5.31 & 981 & 1042 & 628 & 1656 & 6.27 & 157 & 793 & 3.57 & 1925 & 3.22 & 2135 & 3.45 & 4279 & 3.11 & 1427 & 3294 & 2215 & 3401 & 2539 & 3915 & 400 & 4627 \\
Xiaomi Mi 6 & Snapdragon 835 & 133 & 95 & 6.99 & 148 & 6.62 & 825 & 351 & 10.57 & 767 & 11.12 & 680 & 793 & 629 & 1485 & 15.32 & 205 & 523 & 7.97 & 1746 & 9.76 & 1506 & 5.47 & 4456 & 4.16 & 1074 & 3661 & 3106 & 3114 & 1683 & 1936 & 500 & 4621 \\
Samsung Galaxy Note8 & Exynos 8895 Octa & 148 & 84 & 6.73 & 173 & 6.6 & 727 & 467 & 5.19 & 1056 & 5.31 & 655 & 1028 & 628 & 1952 & 6.27 & 714 & 489 & 3.57 & 1839 & 3.22 & 1386 & 3.45 & 4575 & 3.11 & 992 & 2557 & 2129 & 2996 & 1572 & 1522 & 500 & 4555 \\
Samsung Galaxy S8+ & Exynos 8895 Octa & 168 & 69 & 6.73 & 156 & 6.6 & 719 & 408 & 5.19 & 866 & 5.31 & 666 & 1055 & 628 & 1873 & 6.27 & 705 & 470 & 3.57 & 1781 & 3.22 & 1378 & 3.45 & 4457 & 3.11 & 973 & 2497 & 2025 & 3020 & 1477 & 1432 & 400 & 4539 \\
Google Pixel & Snapdragon 821 & 110 & 120 & 6.73 & 163 & 6.6 & 790 & 552 & 5.19 & 912 & 5.31 & 1228 & 1259 & 628 & 2181 & 6.27 & 166 & 874 & 3.57 & 2213 & 3.22 & 2417 & 3.45 & 4663 & 3.11 & 1663 & 3865 & 2146 & 2148 & 1593 & 1464 & 500 & 4538 \\
Samsung Galaxy S8 & Exynos 8895 Octa & 153 & 74 & 6.73 & 165 & 6.6 & 704 & 433 & 5.19 & 970 & 5.31 & 649 & 1084 & 628 & 2078 & 6.27 & 703 & 477 & 3.57 & 1897 & 3.22 & 1394 & 3.45 & 4827 & 3.11 & 987 & 2560 & 2108 & 3275 & 1628 & 1617 & 500 & 4480 \\
OnePlus 5T & Snapdragon 835 & 134 & 421 & 6.73 & 434 & 6.6 & 974 & 1280 & 5.19 & 3108 & 5.31 & 611 & 2458 & 628 & 5814 & 6.27 & 183 & 655 & 3.57 & 2017 & 3.22 & 1722 & 3.45 & 6028 & 3.11 & 1020 & 3338 & 2647 & 5251 & 2182 & 1825 & 500 & 4280 \\
OnePlus 3T & Snapdragon 821 & 117 & 76 & 6.99 & 97 & 6.62 & 902 & 509 & 10.57 & 922 & 11.13 & 1188 & 1285 & 629 & 2139 & 15.32 & 187 & 1092 & 7.97 & 2177 & 9.76 & 2731 & 5.47 & 5336 & 4.16 & 1887 & 4157 & 2785 & 2706 & 1820 & 1755 & 500 & 4122 \\
Sony Xperia XZ1 & Snapdragon 835 & 137 & 397 & 6.99 & 621 & 6.62 & 772 & 1338 & 10.57 & 7138 & 11.13 & 649 & 2610 & 629 & 8164 & 15.32 & 190 & 683 & 7.97 & 2002 & 9.76 & 1763 & 5.47 & 5862 & 4.16 & 1067 & 3490 & 2810 & 4476 & 2203 & 1845 & 400 & 4020 \\
Sony Xperia XZ Premium & Snapdragon 835 & 127 & 480 & 6.99 & 892 & 6.62 & 793 & 1475 & 10.57 & 7865 & 11.13 & 663 & 2458 & 629 & 9751 & 15.32 & 189 & 717 & 7.97 & 1969 & 9.76 & 1750 & 5.47 & 5818 & 4.16 & 1355 & 3742 & 2786 & 3836 & 1644 & 1555 & 500 & 4013 \\
Motorola One Power & Snapdragon 636 & 190 & 100 & 6.73 & 176 & 6.6 & 983 & 411 & 5.19 & 941 & 5.31 & 798 & 893 & 628 & 1910 & 6.27 & 231 & 721 & 3.57 & 2134 & 3.22 & 2088 & 3.45 & 4798 & 3.11 & 1266 & 4030 & 3394 & 3140 & 1599 & 1614 & 400 & 3962 \\
Motorola G7 Plus & Snapdragon 636 & 190 & 110 & 6.73 & 183 & 6.6 & 977 & 423 & 5.19 & 943 & 5.31 & 827 & 920 & 628 & 1949 & 6.27 & 232 & 723 & 3.57 & 2064 & 3.22 & 2152 & 3.45 & 5208 & 3.11 & 1314 & 4141 & 3465 & 3359 & 1670 & 1616 & 400 & 3942 \\
Huawei Honor 8X & Hisilicon Kirin 710 & 163 & 128 & 6.73 & 202 & 6.6 & 1711 & 475 & 5.19 & 962 & 5.31 & 766 & 1008 & 628 & 1998 & 6.27 & 224 & 690 & 3.57 & 2008 & 3.22 & 2148 & 3.45 & 4730 & 3.11 & 1358 & 4338 & 3136 & 3422 & 1615 & 1417 & 400 & 3858 \\
Huawei P smart & Hisilicon Kirin 710 & 164 & 119 & 6.73 & 185 & 6.6 & 1735 & 467 & 5.19 & 946 & 5.31 & 775 & 1016 & 628 & 2163 & 6.27 & 226 & 713 & 3.57 & 2137 & 3.22 & 2094 & 3.45 & 4646 & 3.11 & 1305 & 4267 & 3190 & 3157 & 1615 & 1367 & 400 & 3813 \\
Honor 10 Lite & Hisilicon Kirin 710 & 164 & 126 & 6.73 & 190 & 6.6 & 1701 & 456 & 5.19 & 946 & 5.31 & 771 & 1020 & 628 & 1970 & 6.27 & 229 & 673 & 3.57 & 1980 & 3.22 & 2097 & 3.45 & 4682 & 3.11 & 1310 & 4269 & 3239 & 3189 & 1631 & 1527 & 400 & 3811 \\
Xiaomi Redmi Note 7 & Snapdragon 660 & 190 & 252 & 6.99 & 393 & 6.62 & 904 & 770 & 10.57 & 2481 & 11.13 & 822 & 1485 & 629 & 4084 & 15.32 & 213 & 658 & 7.97 & 2038 & 9.76 & 1794 & 5.47 & 5170 & 4.16 & 1715 & 3700 & 3116 & 4795 & 1980 & 1783 & 400 & 3769 \\
Xiaomi Mi 8 Lite & Snapdragon 660 & 187 & 445 & 6.99 & 884 & 6.62 & 863 & 1291 & 10.57 & 6247 & 11.13 & 700 & 2324 & 629 & 8084 & 15.32 & 207 & 722 & 7.97 & 2208 & 9.76 & 1981 & 5.47 & 6074 & 4.16 & 1052 & 3377 & 3088 & 4898 & 2135 & 2059 & 500 & 3767 \\
Nokia 7 plus & Snapdragon 660 & 188 & 338 & 6.99 & 635 & 6.62 & 865 & 1339 & 10.57 & 5035 & 11.13 & 731 & 2478 & 629 & 6891 & 15.32 & 208 & 765 & 7.97 & 2328 & 9.76 & 2227 & 5.47 & 6119 & 4.16 & 1075 & 3411 & 3187 & 5481 & 1982 & 2072 & 500 & 3746 \\
Samsung Galaxy A9 & Snapdragon 660 & 170 & 466 & 6.73 & 699 & 6.6 & 891 & 1326 & 5.19 & 6055 & 5.31 & 790 & 2516 & 628 & 8895 & 6.27 & 244 & 880 & 3.57 & 2465 & 3.22 & 2251 & 3.45 & 6017 & 3.11 & 1052 & 3354 & 3108 & 5539 & 2121 & 2000 & 500 & 3695 \\
Asus Zenfone 5 & Snapdragon 636 & 133 & 417 & 6.99 & 802 & 6.62 & 860 & 1514 & 10.57 & 6953 & 11.13 & 720 & 2376 & 629 & 8755 & 15.32 & 199 & 801 & 7.97 & 2214 & 9.76 & 2207 & 5.47 & 6087 & 4.16 & 1350 & 4239 & 2836 & 5374 & 2254 & 2133 & 500 & 3686 \\
Nokia X71 & Snapdragon 660 & 188 & 474 & 6.99 & 773 & 6.62 & 939 & 1308 & 10.57 & 6643 & 11.13 & 779 & 2407 & 629 & 9100 & 15.32 & 220 & 843 & 7.97 & 2298 & 9.76 & 2351 & 5.47 & 6254 & 4.16 & 1180 & 3732 & 3340 & 5894 & 2196 & 1998 & 500 & 3537 \\
Xiaomi Mi A2 & Snapdragon 660 & 187 & 326 & 6.99 & 440 & 6.62 & 1214 & 1065 & 10.57 & 1068 & 11.13 & 1217 & 1933 & 629 & 2359 & 15.32 & 208 & 749 & 7.97 & 2187 & 9.76 & 2136 & 5.47 & 6248 & 4.16 & 2107 & 4076 & 3061 & 3919 & 1838 & 1882 & 500 & 3399 \\
Asus Zen. Max Pro M2 & Snapdragon 660 & 188 & 494 & 6.73 & 794 & 6.6 & 960 & 1381 & 5.19 & 5218 & 5.31 & 820 & 2364 & 628 & 8245 & 6.27 & 227 & 905 & 3.57 & 2574 & 3.22 & 2217 & 3.45 & 6892 & 3.11 & 1351 & 4349 & 3544 & 5582 & 2440 & 2181 & 500 & 3377 \\
Samsung Galaxy A7 & Exynos 7885 Octa & 168 & 123 & 6.73 & 231 & 6.6 & 1702 & 535 & 5.19 & 1253 & 5.31 & 1494 & 1176 & 628 & 2407 & 6.27 & 227 & 788 & 3.57 & 2000 & 3.22 & 1966 & 3.45 & 5332 & 3.11 & 1919 & 4938 & 3470 & 4150 & 1803 & 1740 & 500 & 3282 \\
Sony Xperia 10 Plus & Snapdragon 636 & 187 & 447 & 6.99 & 805 & 6.62 & 984 & 1439 & 10.57 & 6274 & 11.13 & 821 & 2496 & 629 & 8640 & 15.32 & 234 & 944 & 7.97 & 2630 & 9.76 & 2447 & 5.47 & 7305 & 4.16 & 1251 & 3977 & 3593 & 5864 & 2721 & 2624 & 400 & 3189 \\
Samsung Galaxy A8 & Exynos 7885 Octa & 167 & 117 & 6.73 & 214 & 6.6 & 1716 & 519 & 5.19 & 1288 & 5.31 & 1455 & 1218 & 628 & 3637 & 6.27 & 227 & 715 & 3.57 & 2294 & 3.22 & 2031 & 3.45 & 5920 & 3.11 & 1916 & 5073 & 3439 & 3395 & 1752 & 1763 & 400 & 3178 \\
Samsung Galaxy A40 & Exynos 7904 & 167 & 106 & 6.73 & 183 & 6.6 & 1727 & 559 & 5.19 & 1152 & 5.31 & 1480 & 1206 & 628 & 2867 & 6.27 & 247 & 800 & 3.57 & 2112 & 3.22 & 2077 & 3.45 & 5636 & 3.11 & 2187 & 5756 & 3969 & 4022 & 2127 & 1943 & 500 & 3127 \\
Samsung Galaxy A30 & Exynos 7904 & 171 & 122 & 6.73 & 219 & 6.6 & 1771 & 497 & 5.19 & 1175 & 5.31 & 1494 & 1158 & 628 & 2637 & 6.27 & 258 & 837 & 3.57 & 2269 & 3.22 & 2071 & 3.45 & 5544 & 3.11 & 2219 & 5113 & 4161 & 4078 & 1925 & 1847 & 500 & 3043 \\
Samsung Galaxy M20 & Exynos 7904 & 170 & 97 & 6.99 & 190 & 6.62 & 1685 & 474 & 10.57 & 1165 & 11.13 & 1475 & 1077 & 629 & 2150 & 15.32 & 249 & 836 & 7.97 & 2164 & 9.76 & 2056 & 5.47 & 5497 & 4.16 & 2238 & 5917 & 4302 & 4334 & 1836 & 1874 & 400 & 2957 \\
Samsung Galaxy A20 & Exynos 7884 & 161 & 110 & 6.73 & 199 & 6.6 & 1787 & 514 & 5.19 & 1203 & 5.31 & 1665 & 1174 & 628 & 2335 & 6.27 & 267 & 842 & 3.57 & 2225 & 3.22 & 2244 & 3.45 & 5688 & 3.11 & 2372 & 6194 & 4396 & 4059 & 2040 & 2199 & 400 & 2892 \\
Huawei P20 lite & HiSilicon Kirin 659 & 209 & 90 & 6.73 & 163 & 6.6 & 1706 & 575 & 5.19 & 1138 & 5.31 & 1045 & 1073 & 628 & 2150 & 6.27 & 497 & 709 & 3.57 & 2645 & 3.22 & 1875 & 3.45 & 5110 & 3.11 & 1564 & 5705 & 5374 & 4237 & 2583 & 2688 & 400 & 2871 \\
Xiaomi Mi A1 & Snapdragon 625 & 230 & 84 & 6.99 & 168 & 6.62 & 1696 & 345 & 10.57 & 918 & 11.13 & 1225 & 807 & 629 & 1837 & 15.32 & 594 & 709 & 7.97 & 2588 & 9.76 & 1648 & 5.47 & 4828 & 4.16 & 1863 & 6160 & 5993 & 2763 & 1851 & 1808 & 500 & 2827 \\
Xiaomi Mi A2 Lite & Snapdragon 625 & 231 & 74 & 6.99 & 140 & 6.62 & 1700 & 346 & 10.57 & 936 & 11.13 & 1204 & 809 & 629 & 1883 & 15.32 & 606 & 761 & 7.97 & 2613 & 9.76 & 1723 & 5.47 & 5047 & 4.16 & 1842 & 6164 & 6026 & 2813 & 1878 & 1941 & 500 & 2795 \\
Motorola Moto G6 Plus & Snapdragon 630 & 234 & 152 & 6.73 & 237 & 6.6 & 1776 & 446 & 5.19 & 1052 & 5.31 & 1146 & 926 & 628 & 2098 & 6.27 & 701 & 828 & 3.57 & 2309 & 3.22 & 2205 & 3.45 & 5374 & 3.11 & 1752 & 5871 & 5553 & 3732 & 1722 & 1875 & 500 & 2790 \\
Motorola P30 Play & Snapdragon 625 & 232 & 98 & 6.73 & 189 & 6.6 & 1690 & 513 & 5.19 & 1197 & 5.31 & 1240 & 828 & 628 & 1915 & 6.27 & 600 & 724 & 3.57 & 2541 & 3.22 & 1785 & 3.45 & 4662 & 3.11 & 1871 & 6243 & 6019 & 2791 & 1690 & 1803 & 400 & 2774 \\
Xiaomi Redmi Note 4 & Snapdragon 625 & 222 & 73 & 6.73 & 146 & 6.6 & 1647 & 357 & 5.19 & 1209 & 5.31 & 1245 & 881 & 628 & 2044 & 6.27 & 592 & 795 & 3.57 & 2846 & 3.22 & 1806 & 3.45 & 5228 & 3.11 & 1873 & 6303 & 6033 & 3106 & 2501 & 2454 & 400 & 2755 \\
Motorola Moto X4 & Snapdragon 630 & 234 & 157 & 6.73 & 230 & 6.6 & 1799 & 437 & 5.19 & 1038 & 5.31 & 1138 & 951 & 628 & 2119 & 6.27 & 706 & 788 & 3.57 & 2299 & 3.22 & 2208 & 3.45 & 5193 & 3.11 & 1714 & 5836 & 5547 & 3330 & 1889 & 1834 & 400 & 2737 \\
Sony Xperia XA2 & Snapdragon 630 & 236 & 185 & 6.99 & 246 & 6.62 & 1846 & 506 & 10.57 & 1186 & 11.13 & 1242 & 1038 & 629 & 2079 & 15.32 & 715 & 924 & 7.97 & 2152 & 9.76 & 2154 & 5.47 & 5667 & 4.16 & 1888 & 6153 & 5605 & 4181 & 1961 & 1894 & 400 & 2519 \\
HTC Desire 19+ & MediaTek Helio P35 & 450 & 161 & 6.77 & 248 & 6.6 & 2049 & 566 & 5.31 & 1146 & 5.2 & 1295 & 1026 & 628 & 2399 & 5.56 & 828 & 1052 & 3.78 & 2947 & 4.31 & 2152 & 3.45 & 6344 & 3.11 & 1662 & 5840 & 6254 & 3387 & 2244 & 2081 & 500 & 2437 \\
Nokia 5 & Snapdragon 430 & 326 & 626 & 6.73 & 846 & 6.6 & 2753 & 1933 & 5.19 & 3469 & 5.31 & 1695 & 3660 & 628 & 5626 & 6.27 & 759 & 1690 & 3.57 & 4447 & 3.22 & 4136 & 3.45 & 10112 & 3.11 & 2568 & 8519 & 9214 & 7110 & 3703 & 3549 & 400 & 1611 \\
\end{tabular}
}
\vspace{2.6mm}
\caption{Benchmark results for several Android devices, a full list is available at:\, \small{\url{http://ai-benchmark.com/ranking}}}
\label{ranking-phones-global}
\vspace{0.2cm}
\end{table*}

\subsection{Floating-point performance}

At the end of September 2018, the best publicly available results for floating-point inference were exhibited by the Kirin 970~\cite{ignatov2018ai}. The increase in the performance of mobile chips that happened here since that time is dramatic: even without taking into account various software optimizations, the speed of the floating-point execution has increased by more than 7.5 times (from 14\% to 100\%, table~\ref{ranking-phones-float}). The Snapdragon 855, HiSilicon Kirin 980, MediaTek Helio P90 and Exynos 9820 launched last autumn have significantly improved the inference runtime for float models and already approached the results of several octa-core Intel CPUs (e.g. Intel Core i7-7700K / i7-4790K) and entry-level Nvidia GPUs, while an even higher performance increase was introduced by the 4th generation of AI accelerators released this summer (present in the Unisoc Tiger T710, HiSilicon Kirin 810 and 990). With such hardware, the Kirin 990 managed to get close to the performance of the GeForce GTX 950~-- a mid-range desktop graphics card from Nvidia launched in 2015, and significantly outperformed one of the current Intel flagships~-- an octa-core Intel Core i7-9700K CPU (Coffee Lake family, working frequencies from 3.60 GHz to 4.90 GHz). This is an important milestone as mobile devices are beginning to offer the performance that is sufficient for running many standard deep learning models, even without any special adaptations or modifications. And while this might not be that noticeable in the case of simple image classification networks (MobileNet-V2 can demonstrate 10+ FPS even on Exynos 8890), it is especially important for various image and video processing models that are usually consuming excessive computational resources.

An interesting topic is to compare the results of GPU- and NPU-based approaches. As one can see, in the third generation of deep learning accelerators (present in the Snapdragon 855, HiSilicon Kirin 980, MediaTek Helio P90 and Exynos 9820 SoCs), they are showing roughly the same performance, while the Snapdragon 855 Plus with an overclocked Adreno 640 GPU is able to outperform the rest of the chipsets by around 10-15\%. However, it is unclear if the same situation will persist in the future: to reach the performance level of the 4th generation NPUs, the speed of AI inference on GPUs should be increased by 2-3 times.
This cannot be easily done without introducing some major changes to their micro-architecture, which will also affect the entire graphics pipeline. It therefore is likely that all major chip vendors will switch to dedicated neural processing units in the next SoC generations.

Accelerating deep learning inference with the mid-range (\eg, Mali-G72 / G52, Adreno 610 / 612) or old-generation (\eg, Mali-T880) GPUs is not very efficient in terms of the resulting speed. Even worse results will be obtained on the entry-level GPUs since they come with additional computational constraints. One should, however, note that the power consumption of GPU inference is usually 2 to 4 times lower than the same on the CPU. Hence this approach might still be advantageous in terms of overall energy efficiency.

One last thing that should be mentioned here is the performance of the default Arm NN OpenCL drivers. Unfortunately, they cannot unleash the full potential of Mali GPUs, which results in atypically high inference times compared to GPUs with a similar GFLOPS performance (\eg the Exynos 9820, 9810 or 8895 with Arm NN OpenCL). By switching to their custom vendor implementation, one can achieve up to 10 times speed-up for many deep learning architectures: \eg the overall performance of the Exynos 9820 with Mali-G76 MP12 rose from 6\% to 26\% when using Samsung's own OpenCL drivers. The same also applies to Snapdragon SoCs which NNAPI drivers are based on Qualcomm's modified OpenCL implementation.

\subsection{Quantized performance}

This year, the performance ranking for quantized inference (table~\ref{ranking-phones-quant}) is led by the Hexagon-powered Qualcomm Snapdragon 855 Plus chipset accompanied by the Unisoc Tiger T710 with a stand-alone NPU. These two SoCs are showing nearly identical results in all int-8 tests, and are slightly (15-20\%) faster than the Kirin 990, Helio P90 and the standard Snapdragon 855. As claimed by Qualcomm, the performance of the Hexagon 690 DSP has approximately doubled over the previous-generation Hexagon 685. The latter, together with its derivatives (Hexagon 686 and 688), is currently present in Qualcomm's mid-range chipsets. One should note that there exist multiple revisions of the Hexagon 685, as well as several versions of its drivers. Hence, the performance of the end devices and SoCs with this DSP might vary quite significantly (\eg, Snapdragon 675 vs. Snapdragon 845).

As mobile GPUs are primarily designed for floating-point computations, accelerating quantized AI models with them is not very efficient in many cases. The best results were achieved by the Exynos 9825 with Mali-G76 MP12 graphics and custom Samsung OpenCL drivers. It showed an overall performance similar to that of the Hexagon 685 DSP (in the Snapdragon 710), though the inference results of both chips are heavily dependent on the running model. Exynos mid-range SoCs with Mali-G72 MP3 GPU were not able to outperform the CPU of the Snapdragon 835 chipset, similar to the Exynos 8890 with Mali-T880 MP12 graphics. An even bigger difference will be observed for the CPUs from the more recent mobile SoCs. As a result, using GPUs for quantized inference on the mid-range and low-end devices might be reasonable only to achieve a higher power efficiency.

%---------------------------------------------------------------------------------------------------------------------------------

\section{Discussion}
\label{sec:discussion}

The tremendous progress in mobile AI hardware since last year~\cite{ignatov2018ai} is undeniable. When compared to the second generation of NPUs (\eg the ones in the Snapdragon 845 and Kirin 970 SoCs), the speed of  floating-point and quantized inference has increased by more than 7.5 and 3.5 times, respectively, bringing the AI capabilities of smartphones to a substantially higher level. All flagship SoCs presented during the past 12 months show a performance equivalent to or higher than that of entry-level CUDA-enabled desktop GPUs and high-end CPUs. The 4th generation of mobile AI silicon yields even better results. This means that in the next two-three years all mid-range and high-end chipsets will get enough power to run the vast majority of standard deep learning models developed by the research community and industry. This, in turn, will result in even more AI projects targeting mobile devices as the main platform for machine learning model deployment.

When it comes to the software stack required for running AI algorithms on smartphones, progress here is evolutionary rather than revolutionary. There is still only one major mobile deep learning library, TensorFlow Lite, providing a reasonably high functionality and ease of deployment of deep learning models on smartphones, while also having a large community of developers. This said, the number of critical bugs and issues introduced in its new versions prevents us from recommending it for any commercial projects or projects dealing with non-standard AI models. The recently presented TensorFlow Lite delegates can be potentially used to overcome the existing issues, and besides that allow the SoC vendors to bring AI acceleration support to devices with outdated or absent NNAPI drivers. We also strongly recommend researchers working on their own AI engines to design them as TFLite delegates, as this is the easiest way to make them available for all TensorFlow developers, as well as to make a direct comparison against the current TFLite's CPU and GPU backends. We hope that more working solutions and mobile libraries will be released in the next year, making the deployment of deep learning models on smartphones a trivial routine.

As before, we plan to publish regular benchmark reports describing the actual state of AI acceleration on mobile devices, as well as changes in the machine learning field and the corresponding adjustments made in the benchmark to reflect them. The latest results obtained with the AI Benchmark and the description of the actual tests is updated monthly on the project website: \url{http://ai-benchmark.com}. Additionally, in case of any technical problems or some additional questions you can always contact the first two authors of this paper.

\section{Conclusions}
\label{sec:conclusion}

In this paper, we discussed the latest advances in the area of machine and deep learning in the Android ecosystem. First, we presented an overview of recently released mobile chipsets that can be potentially used for accelerating the execution of neural networks on smartphones and other portable devices, and provided an overview of the latest changes in the Android machine learning pipeline. We described the changes introduced in the current AI Benchmark release and discussed the results of the floating-point and quantized inference obtained from the chipsets produced by Qualcomm, HiSilicon, Samsung, MediaTek and Unisoc that are providing hardware acceleration for AI inference. We compared the obtained numbers to the results of desktop CPUs and GPUs to understand the relation between these hardware platforms. Finally, we discussed future perspectives of software and hardware development related to this area and gave our recommendations regarding the deployment of deep learning models on smartphones.

{\small
\bibliographystyle{ieee_fullname}
%\bibliography{egbib}

}

\end{document}